\UseRawInputEncoding

\documentclass[aip,prl,amsmath,amssymb,floatfix,reprint,citeautoscript,noeprint,superscriptaddress]{revtex4-2}

\usepackage{bibentry}
\usepackage[english]{babel}
\usepackage{color}
\usepackage{graphicx}
\usepackage[caption=false]{subfig} 
\usepackage{amsmath,amssymb,bm}
\usepackage[version=3]{mhchem}
\usepackage{verbatim}
\usepackage{multirow}
\usepackage{dcolumn}
\usepackage{float}
\usepackage{nicefrac}
\usepackage{siunitx}
\usepackage{booktabs}
\usepackage{chemformula}
\usepackage{wrapfig}
\usepackage{enumitem}  
\usepackage[colorlinks,allcolors=black,citecolor=black,urlcolor=black]{hyperref}
\DeclareSIUnit[number-unit-product = {\,}]{\amu}{amu}
\DeclareSIUnit[number-unit-product = {\,}]{\kJmol}{\kilo\joule\per\mol}
\DeclareSIUnit[number-unit-product = {\,}]{\Nsm}{\newton\second\per\meter\cubed}
\DeclareSIUnit[number-unit-product = {\,}]{\THz}{\tera\hertz}
\DeclareSIUnit[number-unit-product = {\,}]{\meV}{\milli\electronvolt}
\DeclareSIUnit[number-unit-product = {\,}]{\cal}{cal}

\newcommand{\tbf}[1]{\textbf{#1}}

\newcommand{\mbf}[1]{\bm{#1}}
\newcommand{\mrm}[1]{\mathrm{#1}}
\newcommand{\tcr}[1]{\textcolor{black}{#1}}

\newcommand{\etal}{\emph{et al.}}

\begin{document}

\title{Classical quantum friction at water--carbon interfaces}

\author{Anna T. Bui}
\affiliation{Yusuf Hamied Department of Chemistry, University of
  Cambridge, Lensfield Road, Cambridge, CB2 1EW, United Kingdom}

\author{Fabian L. Thiemann}
\affiliation{Thomas Young Centre, London Centre for Nanotechnology,
  and Department of Physics and Astronomy, University College London,
  Gower Street, London, WC1E 6BT, United Kingdom}
\affiliation{Yusuf Hamied Department of Chemistry, University of
  Cambridge, Lensfield Road, Cambridge, CB2 1EW, United Kingdom}
\affiliation{Department of Chemical Engineering, Sargent Centre for
  Process Systems Engineering, Imperial College London, South
  Kensington Campus, London, SW7 2AZ, United Kingdom}

\author{Angelos Michaelides}
\affiliation{Yusuf Hamied Department of Chemistry, University of
  Cambridge, Lensfield Road, Cambridge, CB2 1EW, United Kingdom}

\author{Stephen J. Cox}
\email{sjc236@cam.ac.uk}
\affiliation{Yusuf Hamied Department of Chemistry, University of
  Cambridge, Lensfield Road, Cambridge, CB2 1EW, United Kingdom}

\date{\today}

\begin{abstract}

\tbf{Abstract}: Friction at water--carbon interfaces remains a major puzzle with
theories and simulations unable to explain experimental trends in
nanoscale waterflow. A recent theoretical framework---quantum friction
(QF)---proposes to resolve these experimental observations \tcr{by
considering} nonadiabatic coupling between dielectric fluctuations in
water and graphitic surfaces. Here, using a classical model that
enables fine-tuning of the solid's dielectric spectrum, we provide
evidence from simulations in general support of QF. In
particular, \tcr{as features in the solid's dielectric spectrum} begin
to overlap with water's \tcr{librational and Debye} modes, we find an
increase in friction \tcr{in line with} that proposed by QF. At the
microscopic level, we find that this contribution to friction
manifests more distinctly in the dynamics of the solid's charge
density than that of water. \tcr{Our findings suggest that
experimental signatures of QF may be more pronounced in the solid's
response rather than liquid water's.}

\noindent \tcr{\tbf{Keywords}: liquid--solid friction, nanoscale water, liquid--solid interfaces, graphene, molecular dynamics}
\end{abstract}

\maketitle


Recent advances in nanofluidics \cite{Bocquet2010,Bocquet2020} show
great promise for membrane-based desalination technologies
\cite{Shannon2008,Elimelech2011,Cohen-Tanugi2012} and energy harvesting applications\cite{Simon2008,Zhang2009,Logan2012,Siria2013,Park2014,Siria2017}. 
Owing to the relative ease of fabricating carbon-based nanostructures,
a feature common to many of these technologies is the presence of
extended interfaces between liquid water and carbon.
Despite significant research effort, there are still major gaps
\cite{Bjorneholm2016,Faucher2019,Munoz-Santiburcio2021,Striolo2016}
in our understanding of water at graphitic surfaces.
Of particular curiosity, experiments have found that friction of water
on carbon surfaces is ultra-low compared to other two-dimensional
materials \cite{Holt2006,Whitby2008,Qin2011,Tunuguntla2017,Keerthi2021}.
In addition, friction of water is much higher on multilayer graphite 
\cite{Maali2008,Ortiz-Young2013,Li2016} than monolayer graphene \cite{Xie2018}
and a peculiar radius dependence in multi-walled carbon nanotubes
\cite{Majumder2005, Secchi2016} is observed.
Reproducing these observations has so far remained beyond the realms
of molecular simulations \cite{Falk2010,Thomas2008,Leoni2021,Cicero2008}, even with 
highly accurate interatomic potentials \cite{Thiemann2022}. 
Consequently, these observations cannot be explained by the 
traditional ``surface roughness'' approach
\cite{Bocquet1999,Falk2012} that underpins much of our understanding 
of friction at liquid--solid interfaces.
A recent theoretical study \cite{Kavokine2022} by Kavokine \etal{} has
sought to explain the differences in friction at graphene vs. graphite
by accounting for coupling between collective charge excitations of
the liquid and the dynamics of electrons in the carbon substrate.
In this framework of ``quantum friction'' (QF), friction of water on
graphite is argued to be larger than that on graphene due to the
presence of a dispersionless surface plasmon mode in
graphite\cite{Portail1999,Jensen1991,Xing2021} that overlaps with
liquid water's terahertz (THz) dielectric
fluctuations \cite{Walrafen1990,Ohmine1999,Heyden2010}.
The purpose of the present article is to explore QF with molecular
simulations.

Such coupling between electronic motion in the solid and charge
density fluctuations in the liquid is an effect beyond the
Born--Oppenheimer approximation\cite{Born1927}. 
While simulation schemes to account for such nonadiabatic dynamics
(``electronic friction") exist \cite{Dou2018,Head-Gordon1995,Martinazzo2022}, 
they rely on the accurate construction of a $(3N\times3N)$ friction 
tensor, where $N$ is the total number of atoms explicitly considered 
in the dynamics.
So far, their application has been limited to single gas-phase
molecules on metal
surfaces \cite{Askerka2016,Yin2019,Fuchsel2019,Litman2022}, where the
friction coefficient on each atom can be well-approximated to depend
only on the solid electron density
locally\cite{Hellsing1984,Juaristi2008}.
The low-frequency dielectric modes of water, which are essential to
the description of QF, however, are inherently collective in nature,
prohibiting the application of these sophisticated methods at present.

While accurately accounting for nonadiabatic electronic motion is
computationally challenging, low-frequency dielectric response of water
is reasonably well captured by simple point charge
models \cite{Zarzycki2020,Carlson2020}.
In this article, we therefore focus on this aspect of QF---that
dissipative friction forces are mediated through a complex interplay
of charge density fluctuations---which is more amenable to classical
molecular dynamics (MD) simulations.
By extending a standard treatment for polarizability in graphene such
that its dielectric fluctuations can be precisely controlled, we will
show that coupling between charge density fluctuations in the solid
and liquid increases friction \tcr{in line with the predictions of QF.}
Also similar to QF, this additional contribution is distinct from
the typical surface roughness picture for friction. 
\tcr{The insights afforded by our simulations suggest that 
microscopic signatures of QF manifest more distinctly in the
dynamics of the solid's dielectric fluctuations rather than in 
the structure or dynamics of liquid water.}

\begin{figure*}[t]
    \centering \includegraphics[width=0.98\linewidth]{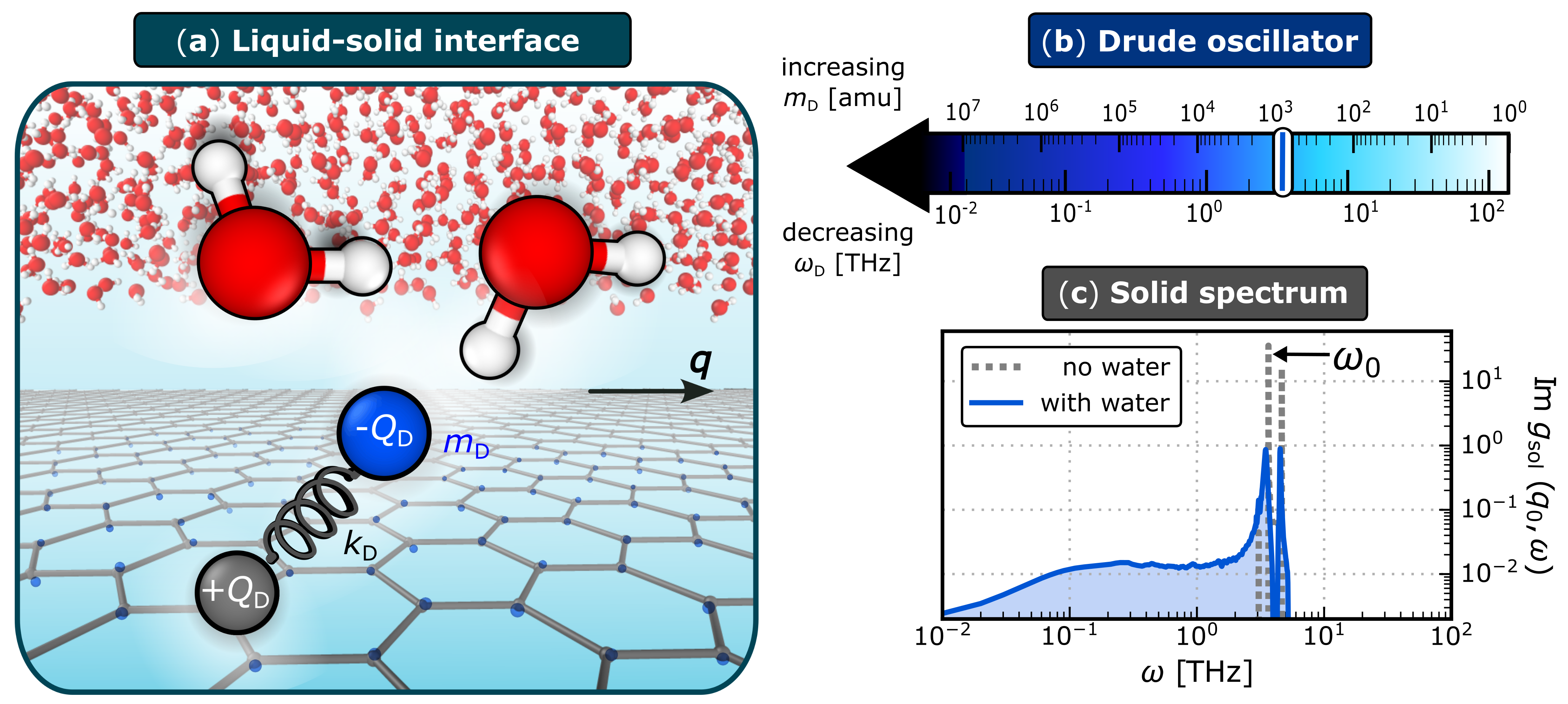} \caption{ \textbf{Model
    of the liquid--solid interface.}  (a) Schematic illustrating the
    interface of a film of water on a graphene sheet. The graphene's
    charge density is described by a classical Drude oscillator model:
    each C atom carries a charge $+Q_{\rm D}$ and is attached to a
    fictitious Drude particle of mass $m_{\mrm{D}}$ and charge
    $-Q_{\rm D}$ via a harmonic spring with force constant
    $k_{\mrm{D}}$. O, H, C atoms and Drude particles are in red,
    white, grey and blue, respectively. For clarity, only one Drude
    particle is highlighted and its displacement from the C atom is
    exaggerated (see SI for the actual distribution).  The
    characteristic frequency of the Drude
    oscillators, \tcr{$\omega_{\rm D} = (k_{\rm D}/m_{\rm D})^{1/2}$}
    is controlled by varying $m_{\mrm{D}}$, as indicated by the
    colorbar in (b). \tcr{(c) In the absence of water, the surface
    response function of the solid, $g_{\rm sol}(q_0,\omega)$ (shown
    for $\omega_{\rm D} = \SI{3.3}{\THz}$), is dominated by two peaks,
    with $\omega_0$ describing the position of the lower frequency
    peak.}  In the present case,
    $\omega_0\approx\SI{3.5}{\THz} \approx \omega_{\rm D}$, indicating
    a weak coupling between the Drude oscillators.  With water
    present, these peaks are broadened.  }
\label{main_model}
\end{figure*}


\textbf{Model of the liquid--solid interface.}
The system we consider consists of a thin film of water on 
a frozen flat graphene sheet, as shown schematically in Fig.~\ref{main_model}(a).
To model water, we use the SPC/E model \cite{Berendsen1987}, which
reasonably captures both the librational modes (hindered molecular rotations)
as a sharp peak at $\omega_{\rm lib}\approx\SI{20}{\THz}$, and the
Debye modes (hindered molecular translations) as a broad feature
spanning $\sim10^{-2}-10^{1}\,\si{\THz}$
\cite{Carlson2020,Zarzycki2020}.
Water--carbon interactions are modeled with a Lennard-Jones potential
that reproduces the contact angle of water droplets on graphitic
surfaces \cite{Werder2003}; while such a potential captures the
essential features of surface roughness contributions to friction, it
lacks any dielectric response.
For each carbon center we therefore also ascribe a charge $+Q_{\rm D}$,
and attach to it, via a harmonic spring with force constant $k_{\rm D}$,
a ``Drude particle'' of mass $m_{\rm D}$ and charge $-Q_{\rm D}$. 
This classical Drude oscillator model is a common approach for modeling
electronic polarizability \cite{Lamoureux2003}, and introduces 
electrostatic interactions between both the water film and the
substrate, and the substrate with itself. 
In the absence of water, the graphene sheet can be considered a set of
weakly interacting harmonic oscillators (see SI).
To parameterize the model, we set $Q_{\rm D} = 1.852\,e$ 
and $k_{\rm D} = \SI{4184}{\kJmol}\mrm{\AA}^{-2}$, 
which have been shown to recover the polarizability tensor of a 
periodic graphene lattice \cite{Misra2017}. 
In usual treatments of electronic polarizability, 
one follows a Car--Parrinello-like scheme \cite{Car1985}
whereby $m_{\rm D}$ is chosen to be sufficiently small to
ensure adiabatic separation of the Drude and nuclear 
(in this case, water) motions. 
Here, we are inspired by the fact that, even for bulk systems,
increasing $m_{\rm D}$ leads to nuclear motion experiencing drag
forces \cite{Sprik1991}.
We therefore treat $m_{\rm D}$ as a free parameter that tunes 
the frequency $\omega_{\rm D}= (k_{\rm D}/m_{\rm D})^{1/2}$ of
an individual oscillator, anticipating that this may lead to
an increase in friction at the liquid--solid interface.
However, the details of how friction may vary with $\omega_{\rm D}$ 
are not \emph{a priori} obvious. 
In practice, we choose $1\lesssim m_{\rm D}/{\rm amu} \lesssim 10^7$, 
such that $10^{-2}\lesssim \omega_{\rm D}/{\rm THz}\lesssim 10^{2}$
as indicated in Fig.~\ref{main_model}(b).
Importantly, changing $m_{\rm D}$ in this manner does not affect the
system's static equilibrium properties. 

In QF, significant overlap between
the substrate and water is due to a dispersionless 
plasmon mode present in graphite but not graphene.
While we cannot reasonably expect the classical Drude
model to faithfully describe this plasmonic behavior,
we can ask a more general question concerning how
friction is affected when the substrate and fluid 
spectra overlap significantly.
This question can readily be addressed by tuning
$m_{\rm D}$, as we describe above, without the need to
introduce multilayer graphite.  
For simplicity, and ease of comparison between
systems, we therefore employ a single graphene sheet 
in all simulations.

\begin{figure*}[t]
    \centering \includegraphics[width=0.98\textwidth]{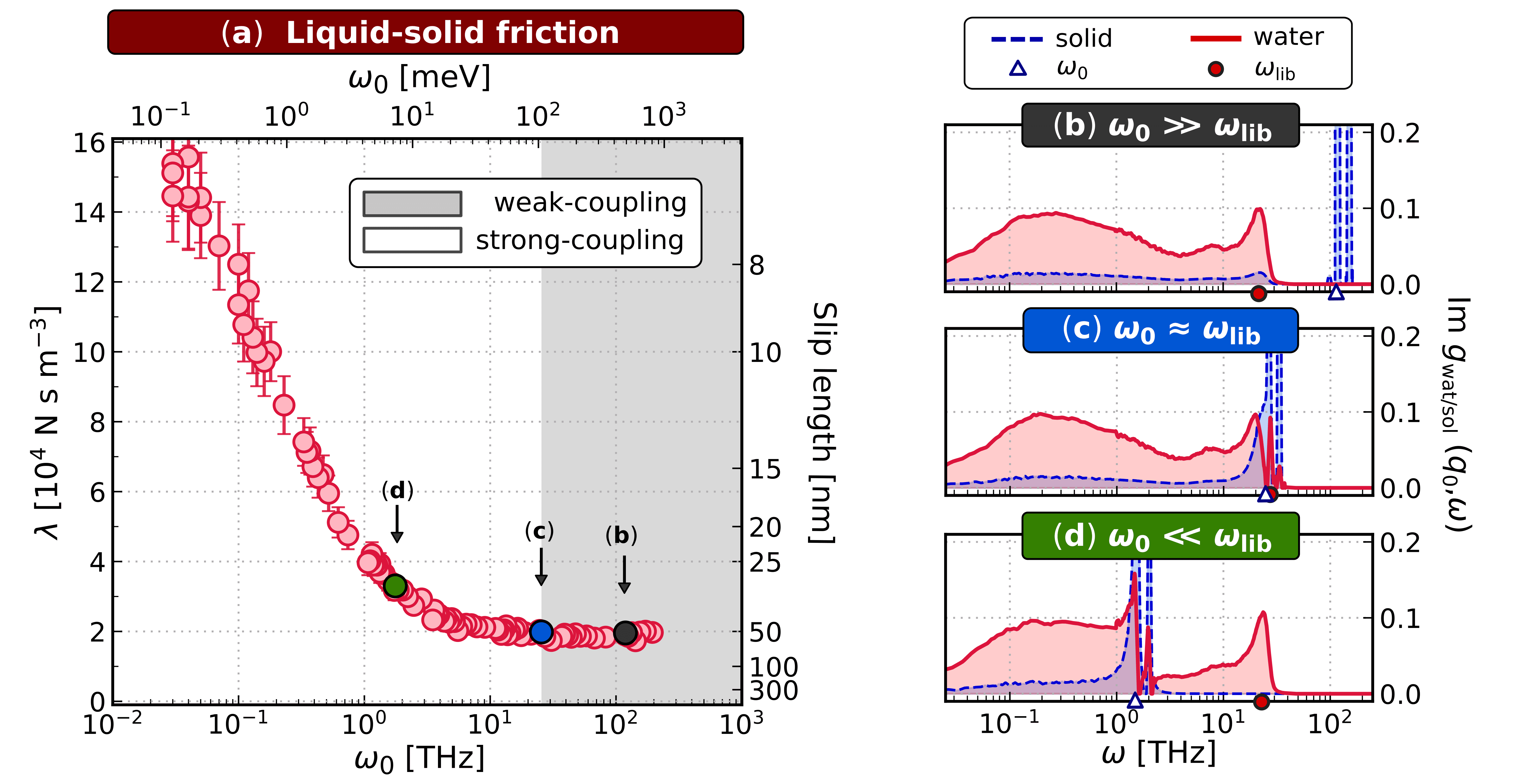} \caption{ \textbf{\tcr{Friction
    increases as dielectric fluctuations \tcr{begin to} overlap.}}
    (a) The liquid--solid friction coefficient $\lambda$ is shown
    against $\omega_0$. The slip length is given as $b=\eta$/$\lambda$
    where $\eta$ is the viscosity of water. Statistical errors are
    obtained from block-averaging.  Two regimes are indicated:
    weak-coupling (shaded gray) where $\lambda$ remains roughly
    constant; and strong-coupling (not shaded) where $\lambda$
    increases with decreasing $\omega_0$.  In (b-d), \tcr{ $g_{\rm
    wat}(q_0, \omega)$ and $g_{\rm sol}(q_0, \omega)$ } are shown for
    three representative cases $\omega_0 \gg \omega_{\rm lib}$,
    $\omega_0\approx \omega_{\rm lib}$ and $\omega_0\ll \omega_{\rm
    lib}$, respectively. We see that the increase in $\lambda$
    coincides with $\omega_0 \lesssim \omega_{\rm lib} \approx
    20$\,THz. \tcr{In addition, when the overlap between the spectra is
    significant, the dominant features of $g_{\rm sol}(q_0,\omega)$
    are broadened, and $g_{\rm wat}(q_0, \omega)$ is perturbed.} The
    boundary between the regimes is approximate.  }
\label{main_friction}
\end{figure*}

Overall, our model describes two fluctuating charge densities,
$n_\mrm{wat}(\mbf{r},t)$ of the water and $n_\mrm{sol}(\mbf{r},t)$
of the solid, originating from the collective motion of water
molecules and Drude oscillators, respectively, at position $\mbf{r}$ 
and time $t$.
The total charges of both the water and the solid are strictly
conserved. 
\tcr{
It will be convenient to characterize these charge
distributions by their surface response functions\cite{Kavokine2022,Pitarke2006}, e.g., for water,
\begin{multline}
\mathrm{Im}\,g_{\mrm{wat}}(q,\omega) = 
\frac{\pi\omega}{q \mathcal{A} k_{\mathrm{B}} T}
\int_{-\infty}^\infty\!\mrm{d}t\,\mrm{e}^{i\omega t}\\
\sum_{\alpha,\beta\,\in\,{\mrm{wat}}}\!\!
\big\langle Q_{\alpha} Q_{\beta}\,
\mrm{e}^{-i\mbf{q}\cdot[\mbf{x}_{\alpha}(t)-\mbf{x}_{\beta}(0)]}
\mrm{e}^{-q |\Delta z_{\alpha}(t)|}\,
\mrm{e}^{-q |\Delta z_{\beta}(0)|}
\big\rangle,
\end{multline}
where $\mathcal{A}$ is the interfacial lateral area, $k_{\rm B}$ is
Boltzmann's constant, $T$ is the temperature, $\mbf{q}$ 
is a wavevector parallel to 
the surface and $Q_\alpha$ is the charge on atom $\alpha$
whose position in the plane of the graphene sheet
at time $t$ is $\mbf{x}_\alpha(t)$
with vertical coordinate $z_\alpha(t)
= z_0 + \Delta z_\alpha(t)$, where
$z_0$ defines a plane between the carbon
atoms and the water contact layer.
The surface response function of the solid,
$g_{\mrm{sol}}(q,\omega)$, is similarly defined. 
}

\tcr{
In Fig.~\ref{main_model}(c), we present $g_{\mrm{sol}}(q_0,\omega)$ for
$m_{\mrm{D}}=10^3\,\si{\amu}$  both in the absence and presence of
water, where $q_0 = 2\pi/L_x \approx \SI{0.25}{\AA^{-1}}$ corresponds 
to a low wavevector accessible in the simulation box.
In the absence of water, $g_{\mrm{sol}}(q_0,\omega)$ exhibits two
dominant peaks (see SI). We will focus on the lower frequency peak,
whose position we take to be $\omega_{\mrm{0}}$. As
$\omega_{\mrm{0}}\approx\omega_{\mrm{D}}$, it is appropriate to
consider the graphene sheet as a set of weakly coupled harmonic
oscillators (see SI). In the presence of water, both of these peaks
are broadened, and we also see the emergence of a broad feature at low
frequencies.} We will discuss the implication of these observations in
the context of friction below. Further technical details of the model,
simulation setup, precise definitions of computed quantities and
additional tests for the sensitivity of our results \tcr{to the choice
of simulation settings} are given in the SI.

\textbf{Friction at the water--carbon interface depends sensitively on $\omega_0$.}
We proceed to explore how the features of \tcr{$g_{\mrm{sol}}(q_0,\omega)$}
affect friction at the interface.
For each value of $m_{\mrm{D}}$, we perform equilibrium MD simulations 
to extract the liquid--solid friction coefficient $\lambda$ from 
the well-established Green--Kubo relationship \cite{Bocquet1999,Bocquet2013}:
\begin{equation}
  \label{eqn:lambdaGK}
  \lambda = \frac{1}{\mathcal{A}k_{\rm B}T}
  \int_0^\infty\!\mrm{d}\tau\,\langle\mathcal{F}(0)\,\mathcal{F}(\tau)\rangle,
\end{equation}
where 
$\mathcal{F}(\tau)$ is the total force acting on the liquid along
a cartesian direction lateral to the graphene sheet at time $\tau$ and 
$\langle\cdots\rangle$ indicates an ensemble average.
In Fig.~\ref{main_friction}(a), we show the dependence of $\lambda$ on
$\omega_0$ in the range $10^{-2}-10^{2}$ \si{\THz} from a
total of 97 simulations.
Overall, as $\omega_0$ decreases, $\lambda$ stays constant until
$\omega_0 \approx \omega_{\rm lib} \approx 20$\,THz, whereupon further
decreasing $\omega_0$ leads to a significant increase in $\lambda$. To
rationalize this observation, we
inspect \tcr{$g_{\mrm{wat}}(q_0,\omega)$}
and \tcr{$g_{\mrm{sol}}(q_0,\omega)$}, as shown in
Figs.~\ref{main_friction}(b--d), for three representative cases. Based
on the relative positions of $\omega_0$ (the principal frequency of
the solid) and $\omega_{\rm lib}$ (liquid water's librational
frequency), we separate the liquid--solid frictional response into two
regimes:
\begin{enumerate}[label=(\roman*)]
\item{Weak-coupling regime: When $\omega_0 \gtrsim \omega_{\rm lib}$, the friction coefficient
remains roughly constant at $\lambda \approx 1.9 \times
10^4\,\si{\Nsm}$. This value agrees well with previous simulations of
water on graphitic
surfaces \cite{Falk2010,Tocci2014,Tocci2020,Thiemann2022,GovindRajan2019,
Poggioli2021}. In this regime, there is a large separation of
timescales between the dielectric modes of water and the substrate.
As a result, there is little overlap between
\tcr{ $g_{\rm sol}(q_0,\omega)$ and $g_{\rm wat}(q_0,\omega)$},
as seen in Fig.~\ref{main_friction}(b), and \tcr{water's} dynamics are
largely unaffected by varying $\omega_0$. The motions of the Drude
oscillators and the water are not strongly coupled.}
\item{Strong-coupling regime: When $\omega_0 \lesssim \omega_{\rm lib}$, hydrodynamic
friction increases as $\omega_0$ decreases, reaching $\lambda \approx
15 \times 10^4\,\si{\Nsm}$ for $\omega_0 \approx \SI{0.03}{\THz}$.
This change in friction of just over one order of magnitude would lead
to a significant change in the corresponding slip length from
$\sim\SI{60}{\nano\meter}$ to $\sim\SI{7}{\nano\meter}$. For
comparison, experiments have reported water slippage in the range of
$0-200$ \si{\nano\meter} on graphene \cite{Xie2018} and
$8-13$ \si{\nano\meter} on
graphite \cite{Maali2008,Li2016,Ortiz-Young2013}. In this regime,
there is no longer a large separation in timescales between the Drude
oscillators and water's dielectric modes. Consequently, as seen in
Figs.~\ref{main_friction}(c) and~(d),
\tcr{ $g_{\rm sol}(q_0,\omega)$ now
overlaps strongly with water's librational and Debye modes, causing
changes in $g_{\rm wat}(q_0,\omega)$ that reflect the dominant
features of $g_{\rm sol}(q_0,\omega)$.} The onset of this regime is
further supported by the \tcr{broadening of the dominant peaks in
$g_\mrm{sol}(q_0,\omega)$} and \tcr{changes} in the spectrum of the
lateral force on the liquid \tcr{(see SI)}. We conclude that the
increase in friction in this strong-coupling regime is indeed due to
coupling of the dielectric modes in the water and the substrate.}
\end{enumerate}

To test the sensitivity of this separation into strong- and
weak-coupling regimes to the details of the system, we have also
performed simulations with different harmonic potentials for the
Drude oscillators \tcr{and a flexible water model} (see SI). 
While differences in the absolute values of $\lambda$ are expected,
and indeed observed, the increase of $\lambda$ for $\omega_0
\lesssim \omega_{\rm lib}$ is robust.

\begin{figure*}[t]
 \centering \includegraphics[width=\textwidth]{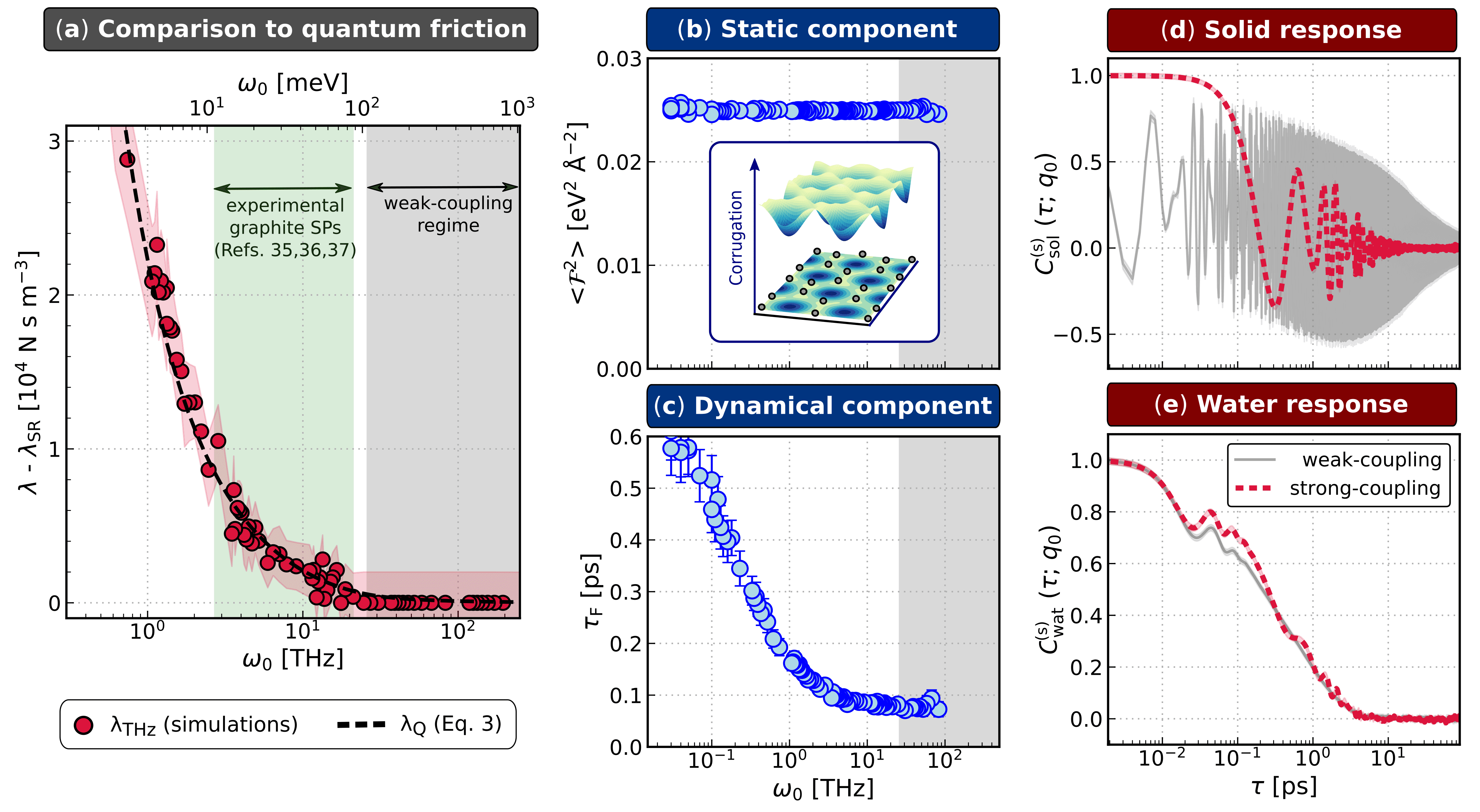} \caption{\textbf{Microscopic
 signatures of quantum friction.} \tcr{ (a) The prediction of QF
 ($\lambda_{\rm Q}$, Eq.~\ref{eqn:lamQF}) well describes $\lambda_{\rm
 THz}$ obtained directly from molecular simulations. The shaded red
 region indicates the standard deviation from block-averaging. A range
 of frequencies for experimental surface plasmons (SPs) in graphite is
 indicated by the green shaded region.}  (b) The static component of
 the total friction $\langle\mathcal{F}^2\rangle$ is essentially
 independent of $\omega_0$ and therefore, all liquid--solid interfaces
 simulated have the same free energy surface shown in the inset (see
 SI for more details).  (c) In contrast, the dynamical component
 $\tau_{\rm F}$ fully captures the dependence of $\lambda$ on
 $\omega_0$.  \tcr{Surface} charge density correlation functions are
 shown for representative cases from the weak-coupling
 ($\omega_0\approx\SI{100}{\THz}$) and
 strong-coupling \tcr{($\omega_0\approx\SI{1}{\THz}$) }regimes for (d)
 the solid and (e) water.  $C^{\mrm{(s)}}_{\rm sol}(\tau; q_0)$ decays
 much more quickly in the strong-coupling regime\tcr{, while changes
 to $C^{\mrm{(s)}}_{\rm wat}(\tau; q_0)$ are much less pronounced.}
 The legend in panel (e) also applies to panel
 (d).}  \label{main_decomposition}
\end{figure*}

\textbf{Comparing molecular simulations with quantum friction theory.}
Before further analysis, it is useful to make a comparison of our
simulation results to QF theory \cite{Kavokine2022}.
Kavokine \etal{} separated the liquid--solid friction into
\(\lambda = \lambda_{\mrm{SR}} + \lambda_{\mrm{Q}}\),
where $\lambda_{\mrm{SR}}$ is the classical surface roughness
contribution and
\tcr{\begin{align}
\lambda_{\rm Q} &= \frac{\hbar^2}{8\pi^2 k_{\rm B}T}\int_0^\infty\!\mrm{d}q\,q^3 \nonumber \\
 &\int_0^\infty
\!\frac{\mrm{d}\omega}{\sinh^2(\hbar\omega/2k_{\rm B}T)}
\frac{\mrm{Im}\,g_{\rm sol}(q,\omega)\,\mrm{Im}\,g_{\rm wat}(q,\omega)}{|1- g_{\rm sol}(q,\omega)\,g_{\rm wat}(q,\omega)|^2} 
\label{eqn:lamQF}
\end{align}}
is the contribution from quantum friction.
In our simulations, changing $m_{\rm D}$ does not affect static
equilibrium properties such as surface roughness (see SI).
In analogy to QF, then, we also decompose the friction coefficient 
from simulation as $\lambda(\omega_0) = \lambda_{\rm SR} +
\lambda_{\rm THz}(\omega_0)$, where $\lambda_{\rm THz}$ 
originates from the coupling of charge density fluctuations in the THz
regime.\footnote{\tcr{In practice, we obtain $\lambda_{\rm SR}$ as the
average value of $\lambda$ in the weak-coupling regime and then obtain
$\lambda_{\rm THz}$ by subtraction.}} 
\tcr{We can obtain approximate
expressions for $g_{\rm sol}(q,\omega)$ and $g_{\rm wat}(q,\omega)$
appropriate for our simulations. As detailed in the SI, for $g_{\rm
wat}(q,\omega)$ we use a parameterization specified in
Ref.~\onlinecite{Kavokine2022} for SPC/E in contact with
graphene/graphite. 
For $g_{\rm sol}(q,\omega)$, we parameterize a
semiclassical Drude model for the surface plasmon\cite{Pisana2007} to roughly capture
the intensity and width of the principal peak of $g_{\rm
sol}(q,\omega)$ observed in our simulations.
In Fig.~\ref{main_decomposition}(a), we compare $\lambda_{\rm Q}$ given by Eq.~\ref{eqn:lamQF} using these suitably
parameterized surface response functions to 
$\lambda_{\rm THz}$ obtained directly from our
simulations. 
The excellent agreement between the simulation result and
Eq.~\ref{eqn:lamQF} provides strong support for the theory of quantum friction outlined in Ref.~\onlinecite{Kavokine2022}.}

\textbf{The microscopic signatures of quantum friction manifest 
in the solid, not the liquid.}
A major advantage of performing molecular simulations is
the insight they can provide at the microscopic scale.
While treating electronic motion as a set of weakly coupled classical
Drude oscillators lacks any explicit treatment of quantum mechanical
effects, the good agreement between this classical model and QF
reinforces the importance of water's low-frequency dielectric modes in
any potential nonadiabatic contributions to friction at water--carbon
interfaces.

Going further, we follow Ref.~\onlinecite{Tocci2020} by disentangling
the origin of the friction at the interface by reformulating
Eq.~\ref{eqn:lambdaGK} as $\lambda
= \langle\mathcal{F}^2\rangle\tau_{\mrm{F}}/(\mathcal{A}k_{\rm B}T)$,
such that the mean-squared force $\langle\mathcal{F}^2\rangle$ and
force decorrelation time $\tau_{\rm F}$ quantify static and
dynamical components, respectively.
As seen in Fig.~\ref{main_decomposition}(b), the static component
remains essentially constant across the entire range of $\omega_0$
explored. \tcr{This implies that the water molecules experience the
same free energy surface at the interface, an example of which is
shown in Fig.~\ref{main_decomposition}(b, inset), independent of
$\omega_0$ (see SI).
This confirms that the physical origins of $\lambda_{\rm THz}$
are not captured by the corrugation of the free energy surface} 
that has been widely used to account for the curvature dependence of
friction in CNTs \cite{Falk2010,Thiemann2022} and certain differences
in hydrodynamic slippage at different materials 
\cite{Ho2011,Falk2012,Tocci2014, Tocci2020,Thiemann2022}.
Instead, the nature of
$\lambda_{\rm THz}$ is entirely dynamical, with the dependence of
$\tau_{\rm F}$ on $\omega_0$ accounting entirely for the
increase in $\lambda_{\rm THz}$, as seen in Fig.~\ref{main_decomposition}(c). 

The above analysis \tcr{demonstrates} that microscopic signatures of
nonadiabatic friction \tcr{should} manifest in dynamical rather than static
properties of the system.
\tcr{We therefore consider the surface charge densities, e.g, for water,
\begin{equation*}
\tilde{n}^{\mrm{(s)}}_\mrm{wat}(q,t) = \sum_{\alpha\in\rm
wat}Q_\alpha \mrm{e}^{i\mbf{q}\cdot\mbf{x}_\alpha(t)} 
\mrm{e}^{-q|\Delta z_{\alpha}(t)|},
\end{equation*}
and inspect their autocorrelation functions
$C^{\mrm{(s)}}_{\mrm{sol}}(\tau; q)$ and
$C^{\mrm{(s)}}_{\mrm{wat}}(\tau; q)$; these are presented in
Figs.~\ref{main_decomposition}(d) and~\ref{main_decomposition}(e)
respectively, for $q=q_0$. While small changes in $C^{\mrm{(s)}}_{\mrm{wat}}(\tau;
q)$ are observed between the weak- and strong-coupling regimes, the
impact on $C^{\mrm{(s)}}_{\mrm{sol}}(\tau; q)$ is much more
pronounced. For the water film, we have also probed molecular
reorientation and hydrogen bond relaxations, and found that these are
barely affected between the two regimes. This suggests that quantum
friction is unlikely to have a significant impact on water's local dynamical
properties.}

We attribute these contrasting behaviors of the liquid and the solid
to the rigidity of water's hydrogen-bond network, which lacks a clear
counterpart from the perspective of the Drude oscillators. In fact, it
is even useful to simply compare the relative magnitude of the dipoles
for a single water molecule $\mu_{\rm wat}$ and a Drude oscillator
$\mu_{\rm D}$. With our simple point charge model, we have
$\langle\mu_{\rm wat}\rangle = 2.351\,\rm{D}$ while $\langle\mu_{\rm
D}\rangle = Q_{\rm D}(3 k_{\rm B}T/k_{\rm D})^{1/2} \approx
0.4\,\rm{D}$. Thus, while the water molecules only feel the presence
of the Drude oscillators as a small perturbation relative to their
intermolecular interactions, the Drude oscillators feel the impact of
the water molecules much more strongly. We speculate that this
conclusion also applies to cases where electronic degrees of freedom
have been accurately accounted for.

In summary, by using a simple model of charge density fluctuations in
a carbon substrate in which we can finely tune \tcr{the surface
response function of the substrate,} we find increases in interfacial
friction in line with those suggested by a recent theory of quantum
friction. We see that the friction increases once \tcr{the principal
peak in the substrate's surface response function overlaps with
features in water's surface response function arising from its
librational and Debye modes.} We show that this extra contribution to
the friction is entirely dynamical in its origin, with static
equilibrium properties apparently indifferent to the degree of
coupling between the water and the substrate.
The insights provided by our molecular simulations reveal that the
increase in friction manifests at the microscopic scale as
a \tcr{pronounced change in the relaxation of the substrate's
dielectric modes, with relatively little impact on the behavior of
water.}

Our model, while able to provide a proof of concept for QF, does not
aim to be a rigorous description of water on graphite. We have
considered a static graphene sheet, which precludes any role that
phonon modes might
play \cite{Ma2015,Bocquet2015,Cruz-Chu2017,Marbach2018,Ambrosetti2022}. Any
changes in surface roughness upon changing from single to multilayer
systems have also not been accounted for. Going forwards, it will be
essential to explore how these factors affect both the surface
roughness and charge density coupling contributions to
friction. Advances in simulations of nonadiabatic
effects\cite{Head-Gordon1995, Mazzola2012,Tubman2014} to accurately
describe the solid's electronic excitations in response to collective
fluctuations in the liquid will also be a welcome development. \tcr{An
obvious limitation of the present model is that it is restricted to
describing the substrate as a dielectric, rather than a conductor (or
semimetal). In principle, extending the current methodology to
classical representations of metallic
substrates \cite{Coretti2022,Siepmann1995} should be relatively
straightforward.}

Despite its simplifications, our model captures the increase in the
interfacial friction when there is an overlap in the dielectric
spectra of the liquid and the solid.
It is important to stress that this principle can be generalized
to the interfaces of any combination of polar liquid and solid.
Since the THz densities of state of a liquid can be reasonably
described in simulations, our model opens up the possibility to
predict whether different liquids \cite{Cazade2014,Futamura2017} 
also show a significant QF component.
In addition to providing early evidence from simulations 
in general support of QF theory, our results suggest a potentially
useful strategy for experimental verification. 
Specifically, the apparent asymmetry between the impact on water and
the substrate suggests it may be advantageous to focus experimental
efforts on spectroscopies that probe the substrate's electronic
response\cite{Ulbricht2011}, rather than seeking hallmarks in the
structure or dynamics of the liquid.

\begin{acknowledgments}
This work was performed using resources provided by the
Cambridge Service for Data Driven Discovery (CSD3) operated by the
University of Cambridge Research Computing Service
(www.csd3.cam.ac.uk), provided by Dell EMC and Intel using Tier-2
funding from the Engineering and Physical Sciences Research Council
(capital grant EP/T022159/1), and DiRAC funding from the Science and
Technology Facilities Council (www.dirac.ac.uk). A.T.B. acknowledges
studentship funding from the Ernest Oppenheimer Fund and Peterhouse
College, University of Cambridge.  S.J.C. is a Royal Society
University Research Fellow (Grant No. URF\textbackslash
R1\textbackslash 211144) at the University of Cambridge. For the
purpose of open access, the authors have applied a Creative
Commons Attribution (CC BY) licence to any Author Accepted Manuscript
version arising.
\end{acknowledgments}

\section*{Supporting Information}

The supporting information provides additional details on the
results presented in the main article. This includes: the model 
and simulation details; precise definitions and computational details 
of the quantities presented in the article;  further analysis
on the coupling of the liquid and solid charge densities;
sensitivity of the results to certain aspects of the simulations 
and the model; detailed comparison to quantum friction theory
and analyses on additional properties of the interface.

\section*{Data availability statement}
The data that support the findings of this study
are openly available at the University of Cambridge Data Repository 
at https://doi.org/10.17863/CAM.89536.

\bibliography{main_bibliography.bib}

\end{document}


\title{Supporting Information for: Classical Quantum Friction at Water--Carbon Interfaces}

\author{Anna T. Bui}
\affiliation{Yusuf Hamied Department of Chemistry, University of
  Cambridge, Lensfield Road, Cambridge, CB2 1EW, United Kingdom}

\author{Fabian L. Thiemann}
\affiliation{Thomas Young Centre, London Centre for Nanotechnology,
  and Department of Physics and Astronomy, University College London,
  Gower Street, London, WC1E 6BT, United Kingdom}
\affiliation{Yusuf Hamied Department of Chemistry, University of
  Cambridge, Lensfield Road, Cambridge, CB2 1EW, United Kingdom}
\affiliation{Department of Chemical Engineering, Sargent Centre for
  Process Systems Engineering, Imperial College London, South
  Kensington Campus, London, SW7 2AZ, United Kingdom}
  
\author{Angelos Michaelides}
\affiliation{Yusuf Hamied Department of Chemistry, University of
  Cambridge, Lensfield Road, Cambridge, CB2 1EW, United Kingdom}

\author{Stephen J. Cox}
\affiliation{Yusuf Hamied Department of Chemistry, University of
  Cambridge, Lensfield Road, Cambridge, CB2 1EW, United Kingdom}
\email{sjc236@cam.ac.uk}

\date{\today}
\maketitle

This supplementary information provides additional details on the
results presented in the main article. This includes: the model 
and simulation details; precise definitions and computational details 
of the quantities presented in the article;  further analysis
on the coupling of the liquid and solid charge densities;
sensitivity of the results to certain aspects of the simulations 
and the model; comparison to quantum friction theory and 
analyses on additional properties of the interface.

\tableofcontents
\newpage


\section{Classical molecular dynamics simulation details}
\label{sec:simulations}

\subsection{Model description}

We consider a system of a film of liquid water on a flat 
graphene sheet as described in the main article. 
Liquid water can be modeled by rigid simple point charge models
with potential energy functions of the form
\begin{equation}
\label{Eq:waterinteraction}
\mcl{U}_{\mrm{wat}}(\mbf{R}_{\mrm{wat}}^{N}) =
 \sum_{i<j}^{N}u_{\mrm{LJ}}
(|\mbf{r}_{\mrm{O},i}-\mbf{r}_{\mrm{O},j}|)
+ \sum_{i<j}^{N}\sum_{\alpha,\beta} 
\frac{Q_{\alpha,i}Q_{\beta,j}}{|\mbf{r}_{\alpha,i}-\mbf{r}_{\beta,j}|},
\end{equation}
where $\mbf{R}_{\mrm{wat}}^{N}$ denotes the set of atomic positions
for a configuration of $N$ water molecules, $\mbf{r}_{\mrm{O},i}$
denotes position of the oxygen atom on water molecule $i$ and
$Q_{\alpha,i}$ is the charge of site $\alpha$ located at position
$\mbf{r}_{\alpha,i}$.
The first set of sums in Eq.~\ref{Eq:waterinteraction}
captures short-ranged repulsion and non-electrostatic 
``long-ranged'' attraction  between water molecules
with the usual Lennard-Jones 12-6 potential 
\begin{equation}
u_{\mrm{LJ}}(r) = 4\epsilon
\left[\left(\frac{\sigma}{r}\right)^{12}
- \left(\frac{\sigma}{r}\right)^{6} \right],
\label{Eq:lj}
\end{equation}
which is parameterized by an energy scale $\epsilon$ and a
length scale $\sigma$.
The second set of sums in Eq.~\ref{Eq:waterinteraction} describes
electrostatic interactions.
Here, we adopt a unit system for electrostatics in
which $4\pi\epsilon_{0}=1$ where $\epsilon_{0}$
is the permittivity of free space.

In the standard surface roughness picture, the liquid interacts
with the solid through long-ranged van der Waals attraction
and short-range Pauli repulsion, which we model with
a 12-6 Lennard-Jones potential between the oxygen atoms
on the water and the carbon atoms in the graphene sheet
\begin{equation}
\mcl{U}_{\mrm{SR}}(\mbf{R}_{\mrm{wat}}^{N},\mbf{R}_{\mrm{sol}}^{M}) =
\sum_{i}^{N}\sum_{j}^{M}u_{\mrm{LJ}}
(|\mbf{r}_{\mrm{O},i}-\mbf{r}_{\mrm{C},j}|),
\end{equation}
where $\mbf{R}_{\mrm{sol}}^{M}$ denotes the set of atomic positions
for a configuration of $M$ carbon atoms.
Here, for simplicity, we fix the positions of the carbon atoms.

Polarization of the solid can then be incorporated with the 
classical Drude oscillator model \cite{Lamoureux2003}.
Each carbon core now carries a positive charge $+Q_{\mrm{D}}$ 
and is attached to a Drude particle of charge $-Q_{\mrm{D}}$ 
and mass $m_{\mrm{D}}$ through a harmonic spring with force
constant $k_{\mrm{D}}$.
The potential energy function of the solid then has the form
\begin{equation}
\mcl{U}_{\mrm{sol}}(\mbf{R}_{\mrm{sol}}^{M}) = \sum_{i}^{M}\frac{1}{2} k_{\mrm{D}}
|\mbf{r}_{\mrm{C},i}-\mbf{r}_{\mrm{D},i}|^{2}
+ \sum_{i<j}^{M}\sum_{\alpha,\beta} 
\frac{Q_{\alpha,i}Q_{\beta,j}}{|\mbf{r}_{\alpha,i}-\mbf{r}_{\beta,j}|}\,
\phi(|\mbf{r}_{\alpha,i}-\mbf{r}_{\beta,j}|).
\end{equation}
The first set of sums describes the total harmonic interaction
of all Drude oscillators in the solid.
The second set of sums describes the electrostatic interactions
between the oscillators. 
To avoid the ``polarization catastrophe'',  a Thole function\cite{Thole1981}, $\phi(r)$,
is used to damp Coulomb interactions at short distances.
This function has the form
\begin{equation}
\phi(r) = 1 - \left( 1+ \frac{s r}{2} \right) e^{-s r},
\label{Eq:thole}
\end{equation}
where the scaling coefficient $s$ is determined by the polarizability 
of the carbon atom $\alpha_{\mrm{C}}$ and
a Thole damping parameter $\delta_{\mrm{C}}$ via $s=\delta_{\mrm{C}}/(\alpha_{\mrm{C}})^{1/3}$.
In the absence of a field, each Drude particle oscillates around
its core atom position $\mbf{r}_{\mrm{C}}$. 
%
In an electric field $\mbf{E}(t)$, here arising 
from the fluctuating charge density of water, the Drude
particle oscillates around $\mbf{r}_{\mrm{C}}$ + $\mbf{d}(t)$ 
where $\mbf{d}(t)$ is the displacement of Drude particle
from the core atom at time $t$ and is given by $\mbf{d}(t)=Q_{\mrm{D}}\mbf{E}(t)/k_{\mrm{D}}$.
The instantaneous induced dipole from the Drude oscillator is 
$\mbf{\mu}_{\mrm{D}}(t)=Q_{\mrm{D}}\mbf{d}(t)=Q_{\mrm{D}}^2\mbf{E}(t)/k_{\mrm{D}}$.
The isotropic atomic polarizability is then seen to be 
\cite{Schroder2010,Lamoureux2003,Noskov2005,Lamoureux2006}
\begin{equation}
\alpha_{\mrm{C}} = \frac{Q_{\mrm{D}}^2}{k_{\mrm{D}}}.
\label{eq:polarizability}
\end{equation}
The polarization response of the solid is therefore
controlled by the parameters $Q_{\mrm{D}}$ and $k_{\mrm{D}}$.

Introduction of point charges in the solid now introduces 
electrostatic interactions between the solid and the liquid,
which we refer to as the charge density coupling term:
\begin{equation}
\mcl{U}_{\mrm{CC}}(\mbf{R}_{\mrm{wat}}^{N},\mbf{R}_{\mrm{sol}}^{M}) =
\sum_{i}^{N}\sum_{j}^{M}
\sum_{\alpha,\beta} 
\frac{Q_{\alpha,i}Q_{\beta,j}}{|\mbf{r}_{\alpha,i}-\mbf{r}_{\beta,j}|}.
\end{equation}
The overall potential energy of the system is:
\begin{equation}
\mcl{U}_{\mrm{tot}}(\mbf{R}_{\mrm{wat}}^{N},\mbf{R}_{\mrm{sol}}^{M})
= \mcl{U}_{\mrm{wat}}(\mbf{R}_{\mrm{wat}}^{N}) 
+ \mcl{U}_{\mrm{SR}}(\mbf{R}_{\mrm{wat}}^{N},\mbf{R}_{\mrm{sol}}^{M}) 
+ \mcl{U}_{\mrm{sol}}(\mbf{R}_{\mrm{sol}}^{M}) 
+ \mcl{U}_{\mrm{CC}}(\mbf{R}_{\mrm{wat}}^{N},\mbf{R}_{\mrm{sol}}^{M}).
\end{equation}
We reiterate that in previous work\cite{Bocquet1999,Falk2012},
classical treatments of liquid--solid interfacial friction 
have focused on the first two terms of the potential. 
Here, our model considers also interactions between
fluctuating charge densities in the liquid and the solid
through the addition of the last two terms.

\subsection{System set-up}

\begin{figure}[H]
    \centering
    \includegraphics[width=0.9\linewidth]{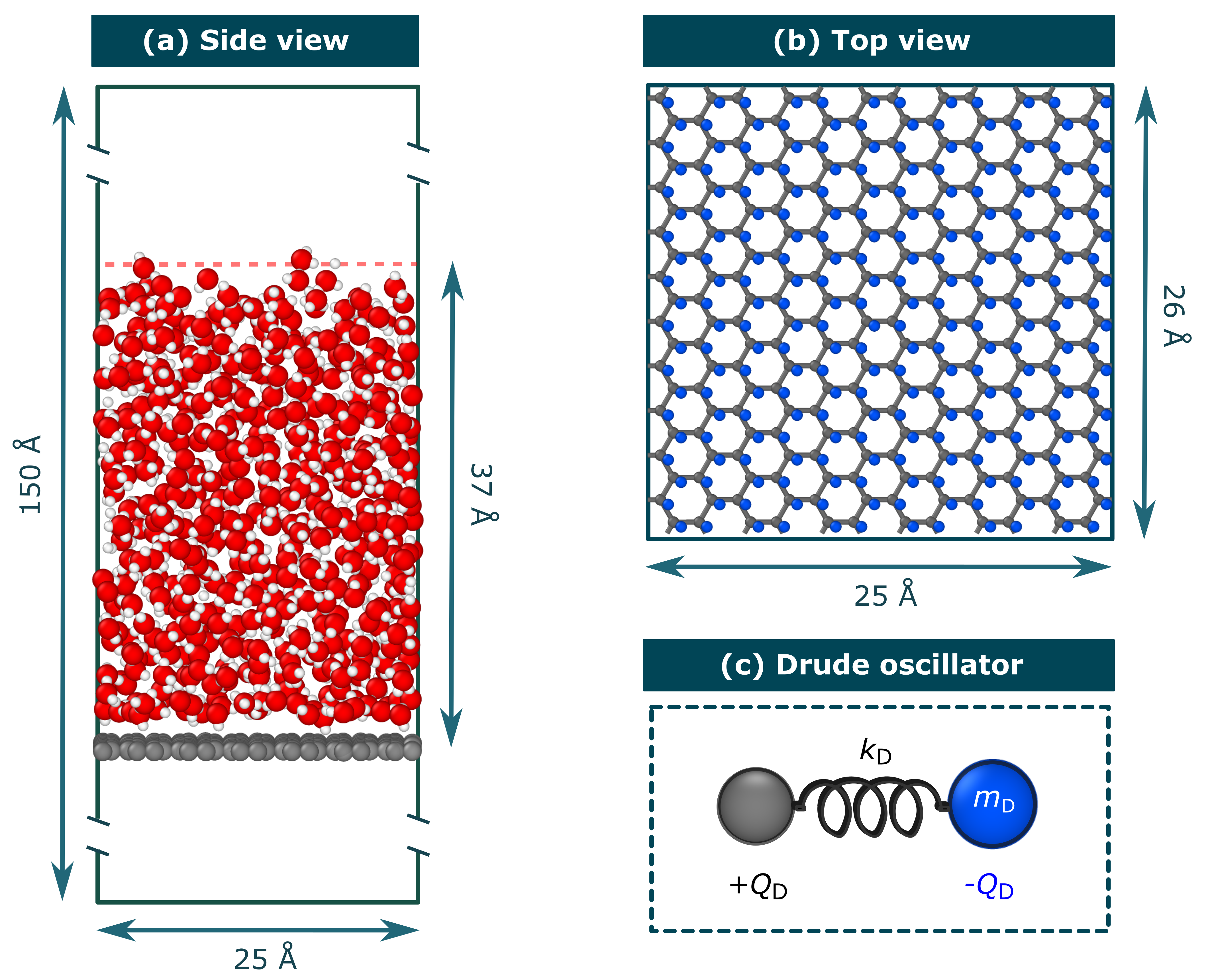}
    \caption{\textbf{System set-up.} The view from (a) the side
    of the liquid--solid interface and (b) the top of the solid sheet.
    Oxygen, hydrogen, carbon atoms and Drude particles are
    in red, white, grey and blue, respectively.
    The dark solid lines represent the edges of the simulation box.
    (c) The classical Drude oscillator is used to model charge 
    density in the solid, whose parameters include: the harmonic
    spring constant $k_{\mrm{D}}$, the Drude mass $m_{\mrm{D}}$,
    charge on the core carbon atom $+Q_{\mrm{D}}$ and on the Drude particle $-Q_{\mrm{D}}$.}
    \label{si_setup}
\end{figure}

For the results presented in the main article,
the simulations were carried out with 724 water molecules
in a thin film of thickness $\approx\SI{37}{\angstrom}$
above a graphene layer of 240 carbon atoms, as illustrated
in Fig.~\ref{si_setup}.
Each carbon atom is attached to one Drude particle so there
are 240 Drude particles.
The orthorhombic cell has dimension 
$\approx26\times25\times\SI{150}{\angstrom\cubed}$.
In addition, simulations of larger system sizes
and of water films with different thicknesses
were also performed to check
the sensitivity of the results, which are presented in Sec.~\ref{sec:sensitivity}.

\subsection{Simulation details}

All simulations were carried out with the \texttt{LAMMPS}
simulations package \cite{Plimpton1995,Thompson2022}. 
Water--water interactions were described with the SPC/E water model
\cite{Berendsen1987}. 
The geometry of water molecules was constrained using the \texttt{RATTLE}
algorithm \cite{Andersen1983}. 
The carbon positions of the sheet were fixed. 
Water--carbon interaction responsible for the surface 
roughness was modeled with Werder parameters 
\cite{Werder2003}. 
Charge densities on the graphene sheet were modeled 
using the classical Drude oscillator model \cite{Lamoureux2003, Dequidt2016} with Thole damping \cite{Thole1981}
using parameters from Misra and Blankschtein\cite{Misra2017}.
The value of parameters for all interaction potentials
in the simulations are summarized in Tabs.~\ref{charges} and \ref{LJparameters}. 
All Lennard-Jones interactions were truncated and shifted at 
\SI{10}{\angstrom}. 
Electrostatic interactions were cut off at \SI{10}{\angstrom} and 
long-ranged interactions were evaluated using particle--particle 
particle--mesh Ewald summation \cite{hockney1988} such that
the RMS error in the forces was a factor of $10^5$ smaller 
than the force between two unit charges separated by a 
distance of \SI{1.0}{\angstrom} \cite{Kolafa1992}.
Drude mass choices to change the solid charge density frequency
are in the range of $1-10^7\,\mrm{amu}$.

\begin{table}[H]
    \centering
    \begin{tabular}{c c c c}
    \hline
    \hline
    Subsystem & Atom type & Mass [amu] & Charge [$\mrm{e}$]  \\
    \hline
    water     &   O       & $15.9994$    &  $-0.8476$               \\
              &   H       & $1.008$      &  $+0.4238$               \\
    \hline
    carbon    &   C       &    *       &  $+1.8520$               \\
              &   D       & $1-10^7$   &  $-1.8520$               \\
    \hline
    \hline
    \end{tabular}
\caption
{\textbf{Parameters for masses and charges for atoms in simulations.}
Parameters for water are from the SPC/E model \cite{Berendsen1987}
and those for the carbon solid are from
Misra and Blankschtein\cite{Misra2017}.
The * denotes that C atoms are fixed so their
masses do not contribute to the dynamics.
The letter D is used to denote Drude particles.}
\label{charges}
\end{table}

\begin{table}[H]
    \centering
    \begin{tabular}{c c l}
    \hline
    \hline
    Interaction potential      & Potential form &  Parameters   \\
    \hline
    SPC/E\cite{Berendsen1987}  & rigid bond  & $r_{\mrm{OH}}=\SI{1.0}{\angstrom}$   \\
                               & rigid angle & $\theta_{\mrm{HOH}}=\ang{109.47}$    \\
                               & Lennard-Jones  & $\epsilon_{\mrm{OO}}=\SI{0.1553}{\kilo\cal\per\mol}$ \\
                               &                & $\sigma_{\mrm{OO}}=\SI{3.166}{\angstrom}$ \\
    \hline
    Werder\cite{Werder2003}    & Lennard-Jones  & $\epsilon_{\mrm{CO}}=\SI{0.1553}{\kilo\cal\per\mol}$ \\
                               &                & $\sigma_{\mrm{CO}}=\SI{3.190}{\angstrom}$ \\
    \hline
    Misra and Blankschtein\cite{Misra2017} & Thole damping\cite{Thole1981}  & $\delta_{\mrm{C}}=1.507$\\
                               & harmonic bond & $k_{\mrm{D}}=\SI{1000}{\kilo\cal\per\mol\per\angstrom\squared}$ \\ 
    \hline
    \hline
    \end{tabular}
\caption
{\textbf{Parameters for interaction potentials (force fields) employed in simulations.} 
For polarization in the solid, the choice
of $k_{\mrm{D}}$ and $Q_{\mrm{D}}$ from Misra and Blankschtein\cite{Misra2017} 
gives an isotropic polarizability of $\alpha_{\mrm{C}}=\SI{1.139}{\angstrom\cubed}$.}
\label{LJparameters}
\end{table}

The simulations were carried out in the canonical (NVT) ensemble, 
where the temperature was held at $\SI{300}{K}$.
Two separate Nos\'{e}--Hoover thermostats \cite{Shinoda2004,Tuckerman2006} 
were applied to the water and Drude particles.
Each thermostat is a Nos\'{e}--Hoover chain with 10 thermostats and a 
damping constant of $\SI{0.1}{\pico\second}$.
Dynamics were propagated using the velocity Verlet algorithm with a 
time-step of $\SI{1}{\femto\second}$, unless specified otherwise. 
Each system was equilibrated for $\SI{100}{\pico\second}$ and the 
subsequent $\SI{10}{\nano\second}$ was used for analysis to give
the results presented in the main article.
The sensitivity of the results to different simulation settings is presented in Sec.~\ref{sec:sensitivity}.

\newpage
\section{Computation of properties}
\subsection{Friction coefficient}
For each equilibrium MD simulation, the friction coefficient was
evaluated through the Green--Kubo formula\cite{Bocquet1996}
involving the time integral of the force autocorrelation function 
defined as:
\begin{equation}
  \label{Eq:lambdaGK}
  \lambda_{\mrm{GK}}(\tau) = \frac{1}{\mcl{A}k_{\rm B}T}
  \int_0^\tau\!\mrm{d}t\,\langle \mcl{F}(0)\cdot \mcl{F}(t)\rangle,
\end{equation}
where $\mathcal{A}$ is the interfacial lateral area,
$\langle\cdots\rangle$ indicates an ensemble average
and $\mcl{F}(t)$ denotes the instantaneous lateral 
force exerted on the liquid by the solid at time $t$.
$\mcl{F}(t)$ is evaluated as the total
summed force acting on all water molecules of a given 
configuration averaged over both in-plane dimensions $(x,y)$
and is saved at every time-step ($\SI{1}{\femto\second}$).
In principle, the friction coefficient of the system is
recovered at the long-time limit:
\begin{equation} 
\lambda = \lim_{\tau\to\infty} \lambda_{\mrm{GK}}(\tau).
\end{equation}
However, at long times, the integral in Eq.~\ref{Eq:lambdaGK} decays
to zero due to the finite lateral extent of the system 
\cite{Bocquet1997,Espanol2019} so evaluating  
$\lambda_{\mrm{GK}}(\tau)$ to a plateau is commonly employed
\cite{Falk2012,Tocci2014,Poggioli2021}.
It has been shown that taking the maximum of $\lambda_{\mrm{GK}}(\tau)$ 
only recovers friction correctly when there is 
a separation of timescales between the decay time
and the memory time of the force autocorrelation function \cite{Oga2019}.
Since we are probing behaviors of the interface
where there is no separation
of timescales, we approximate $\lambda$ as the plateaued
friction coefficient averaged between correlation time 
of $5-10\,\si{\pico\second}$.
Justification of this choice is detailed in Sec.~\ref{sec:sensitivity}.
For each Drude mass, we perform equilibrium MD simulations
to extract $\lambda$. 
The error bars correspond to the statistical errors obtained
from splitting the entire trajectory into 100 blocks
such that each block is $100\,\mrm{ps}$ long.

In the main text, we decompose the static and dynamical
components of the friction coefficient by reformulating 
the Green--Kubo expression in terms of the mean
square force $\langle \mcl{F}^2\rangle$ and the force
decorrelation time $\tau_{\mrm{F}}$:
\begin{equation} 
\lambda = \frac{1}{\mathcal{A}k_{\rm B}T}\langle \mcl{F}^2\rangle\tau_{\mrm{F}},
\end{equation} 
where 
\begin{equation} 
\tau_{\mrm{F}} = \int_0^\infty\!\mrm{d}t\,\frac{\langle \mcl{F}(0)\mcl{F}(t)\rangle}{\langle \mcl{F}^2\rangle}.
\label{eq:tauF}
\end{equation} 
In practice, we computed the friction coefficient and 
the mean square force first before obtaining the force 
decorrelation time via 
$\tau_{\mrm{F}}=\lambda\mathcal{A}k_{\rm B}T / \langle \mcl{F}^2\rangle$.

\subsection{Surface response function}

In the main article, the charge density distributions of the
solid and the liquid are characterized by their surface response functions
defined as
\begin{equation}
\mathrm{Im}\,g_{\mrm{sol}}(q,\omega) = 
\frac{\pi\omega}{q \mathcal{A} k_{\mathrm{B}} T}
\int_{-\infty}^\infty\!\mrm{d}t\,\mrm{e}^{i\omega t}
\sum_{\alpha,\beta\,\in\,{\mrm{sol}}}\!\!
\big\langle Q_{\alpha} Q_{\beta}\,
\mrm{e}^{-i\mbf{q}\cdot[\mbf{x}_{\alpha}(t)-\mbf{x}_{\beta}(0)]}
\mrm{e}^{-q |z_{\alpha}(t)-z_0|}\,
\mrm{e}^{-q |z_{\beta}(0)-z_0|}
\big\rangle,
\label{eq:S_sol}
\end{equation}
and
\begin{equation}
\mathrm{Im}\,g_{\mrm{wat}}(q,\omega) = 
\frac{\pi\omega}{q \mathcal{A} k_{\mathrm{B}} T}
\int_{-\infty}^\infty\!\mrm{d}t\,\mrm{e}^{i\omega t}
\sum_{\alpha,\beta\,\in\,{\mrm{wat}}}\!\!
\big\langle Q_{\alpha} Q_{\beta}\,
\mrm{e}^{-i\mbf{q}\cdot[\mbf{x}_{\alpha}(t)-\mbf{x}_{\beta}(0)]}
\mrm{e}^{-q |z_{\alpha}(t)-z_0|}\,
\mrm{e}^{-q |z_{\beta}(0)-z_0|}
\big\rangle,
\label{eq:S_wat}
\end{equation}
respectively.  Here, $Q_{\alpha}$ is the charge on atom $\alpha$,
whose position in the plane of the graphene sheet at time $t$ is
$\mbf{x}_\alpha(t)$, $\mbf{q}$ is a wavevector parallel to the
graphene sheet, $z_\alpha(t)$ is the vertical coordinate and $z_0 =
1.6$\,\AA{} above the graphene sheet 
defines a plane between carbon atoms and the water contact
layer.

In practice, we computed at every time-step (\SI{1}{\femto\second})
the Fourier--Laplace surface components of the charge densities for the
solid and the liquid, defined as
\begin{equation}
\tilde{n}^{\mrm{(s)}}_{\mrm{sol}}(q,t) = \sum_{\alpha\,\in\,\mrm{sol}}
Q_{\alpha}\mrm{e}^{i\mbf{q}\cdot\mbf{x}_{\alpha}(t)}
\mrm{e}^{-q|z_{\alpha}(t)-z_0|},
\end{equation}
and
\begin{equation}
\tilde{n}^{\mrm{(s)}}_{\mrm{wat}}(q,t) = \sum_{\alpha\,\in\,\mrm{wat}}
Q_{\alpha}\mrm{e}^{i\mbf{q}\cdot\mbf{x}_{\alpha}(t)}
\mrm{e}^{-q|z_{\alpha}(t)-z_0|},
\end{equation}
respectively. As we are interested in the long-wavelength limit
($q \rightarrow 0$), we focus on $\mbf{q}=\mbf{q}_0$, the lowest
wavevector in the $x$ direction accessible in our simulation box, the
magnitude of which is $q_0 =
2\pi/L_x \approx \SI{0.25}{\per\angstrom}$ where $L_x$ is the length
of the box in the $x$ direction.
The power spectra of the surface charge densities are given as
\begin{equation}
S_{\mrm{sol}}^{\mrm{(s)}}(q,\omega)=\frac{1}{\mcl{A}}\int^{+\infty}_{-\infty}\!
\mrm{d}t \,\langle \tilde{n}_{\mrm{sol}}^{\mrm{(s)}}(q,0)
\,\tilde{n}_{\mrm{sol}}^{\mrm{(s)}}(-q,t)\rangle \,
\mrm{e}^{i\omega t},
\end{equation}
and
\begin{equation}
S_{\mrm{wat}}^{\mrm{(s)}}(q,\omega)=\frac{1}{\mcl{A}}\int^{+\infty}_{-\infty}\!
\mrm{d}t \,\langle \tilde{n}_{\mrm{wat}}^{\mrm{(s)}}(q,0)
\,\tilde{n}_{\mrm{wat}}^{\mrm{(s)}}(-q,t)\rangle \,
\mrm{e}^{i\omega t}.
\end{equation}
Through the fluctuation-dissipation theorem, we can obtain the
imaginary part of the surface response function through:
\begin{equation}
\mrm{Im}\,g_{\rm sol}(q,\omega) = 
\frac{2\pi}{q}\frac{\omega}{2k_{\rm B}T}S_{\mrm{sol}}^{\mrm{(s)}}(q,\omega),
\end{equation}
and
\begin{equation}
\mrm{Im}\,g_{\rm wat}(q,\omega) = 
\frac{2\pi}{q}\frac{\omega}{2k_{\rm B}T}S_{\mrm{wat}}^{\mrm{(s)}}(q,\omega).
\end{equation}
To ensure the spectrum is independent of noise, a Savitzky--Golay 
filter \cite{Savitzky1964} was applied.
The resulting spectra without further fitting are presented in the
main article.

\subsection{Surface charge density autocorrelation function}

To characterize relaxation of the solid and the liquid charge densities, 
we computed their respective normalized autocorrelation functions,
defined as:
\begin{equation}
  C^{\mrm{(s)}}_{\mrm{sol}}(\tau; q) = \frac{\langle\tilde{n}^{\mrm{(s)}}_{\mrm{sol}}(q,0)
  \,\tilde{n}^{\mrm{(s)}}_{\mrm{sol}}(-q,\tau)\rangle}
  {\langle|\tilde{n}^{\mrm{(s)}}_{\mrm{sol}}(q)|^2\rangle},
\end{equation}
and
\begin{equation}
  C^{\mrm{(s)}}_{\mrm{wat}}(\tau; q) = \frac{\langle\tilde{n}^{\mrm{(s)}}_{\mrm{wat}}(q,0)
  \,\tilde{n}^{\mrm{(s)}}_{\mrm{wat}}(-q,\tau)\rangle}
  {\langle|\tilde{n}^{\mrm{(s)}}_{\mrm{wat}}(q)|^2\rangle}.
\end{equation}
Again, focusing on the long-wavelength limit, we
show the results for $C^{\mrm{(s)}}_{\mrm{sol}}(\tau; q_0)$
and $C^{\mrm{(s)}}_{\mrm{wat}}(\tau; q_0)$ in the main article.

\newpage
\section{Sensitivity of the friction coefficient}
\label{sec:sensitivity}
To assess the robustness of our results to the choice of simulation
settings, here we present an extensive set of tests on the
sensitivity of the friction coefficients computed to
certain aspects of our simulations. 
In these tests, we show the results for two representative
cases: $m_{\mrm{D}}=1\,\mrm{amu}$ for the weak-coupling regime 
and $m_{\mrm{D}}=5000\,\mrm{amu}$ for the strong-coupling regime.

\subsection{Convergence of the Green--Kubo friction coefficient}

\begin{figure}[H]
    \centering
    \includegraphics[width=\linewidth]{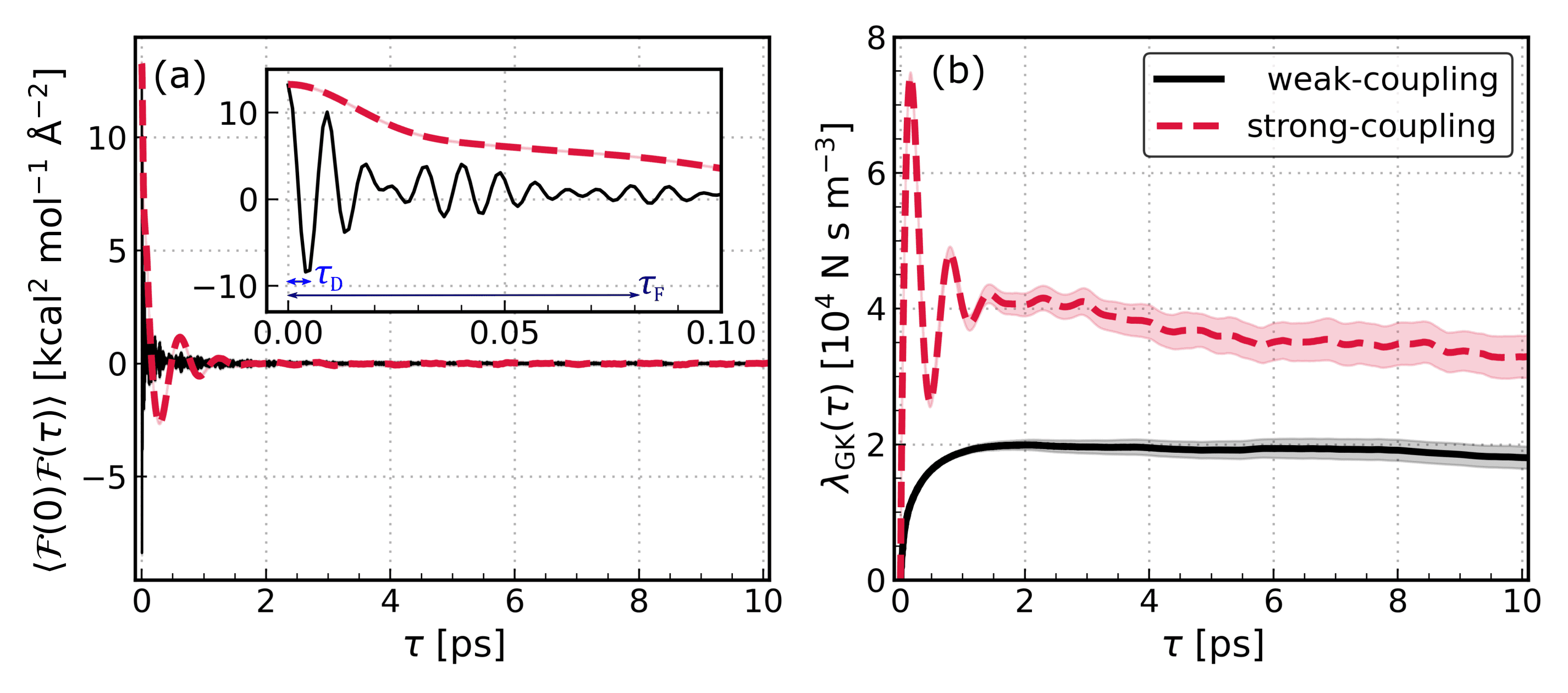}
    \caption{\textbf{Convergence of the Green--Kubo friction coefficient}
    (a) The force autocorrelation function of two 
    representative cases in the weak-coupling and strong-coupling
    regimes. The inset shows clearer the oscillations
    due to the Drude dynamics. The memory time $\tau_{\mrm{D}}$
    and the decay time $\tau_{\mrm{F}}$ are marked for the 
    weak-coupling case.
    (b) The convergence of $\lambda_{\mrm{GK}}$ where statistical 
    errors (shaded area) are obtained from block-averaging.}
    \label{si:friction_convergence}
\end{figure}

As seen from Fig.~\ref{si:friction_convergence},
the force autocorrelation 
$\langle \mcl{F}(0) \mcl{F}(\tau)\rangle$ 
shows oscillations due to charge 
density fluctuations in the solid at short timescales before 
decaying to zero at longer timescales.
We can characterize two particular timescales: (i) 
the time at which the first minimum is reached, $\tau_{\mrm{D}}\propto\omega_{\mrm{D}}$,
due to motion of the Drude oscillators and (ii) the
force decorrelation time, $\tau_{\mrm{F}}\propto\lambda$,
as defined by Eq.~\ref{eq:tauF}.
These are marked for the weak-coupling case in
the inset of Fig.~\ref{si:friction_convergence}(a). 
In $\lambda_{\mrm{GK}}(\tau)$, these timescales manifest in a
peak at $\tau_{\mrm{D}}$ and a plateau after $\tau_{\mrm{F}}$.
In the weak-coupling regime where the Drude mass is low,
there is a separation of timescales as 
$\tau_{\mrm{D}} \ll \tau_{\mrm{F}}$.
A decrease in the solid charge density frequency in this regime 
only increases $\tau_{\mrm{D}}$ but not $\tau_{\mrm{F}}$. 
The peak at $\tau_{\mrm{D}}$ is a local maximum, increasing 
in height as Drude mass increases, and a plateau is observed 
at times intermediate between $\tau_{\mrm{D}}$ and $\tau_{\mrm{F}}$,
unchanged in height. 
This is consistent with the converged friction coefficient
remaining at  $\lambda \approx 1.9 \times 10^4\,\si{\Nsm}$.
In the strong-coupling regime where the solid frequency is
slow, $\tau_{\mrm{D}}$  is now comparable to $\tau_{\mrm{F}}$.
A decrease in the solid charge density frequency in this 
regime increases both $\tau_{\mrm{D}}$ and $\tau_{\mrm{F}}$. 
As $\tau_{\mrm{F}}$ governs the value of $\lambda_{\mrm{GK}}$ 
at long times, the converged friction coefficient now increases
with decreasing charge density frequency.
In order to get a consistent value for the friction coefficient,
it is important to extract $\lambda$ from $\lambda_{\mrm{GK}}(\tau)$
at $\tau > \tau_{\mrm{F}}$. 
Therefore, taking the maximum of $\lambda_{\mrm{GK}}(\tau)$ as
in some studies \cite{Poggioli2021,Seal2021,Joly2016} is not appropriate, 
as also shown previously by Oga \etal{}\cite{Oga2019}

For simulations with the largest Drude mass 
$m_{\mrm{D}}=10^7\,\mrm{amu}$, we observe a plateau in 
$\lambda_{\mrm{GK}}(\tau)$ from $\tau \gtrsim \SI{5}{\pico\second}$.
Therefore, we give the converged friction coefficient 
in all cases to be the average of values of $\lambda_{\mrm{GK}}(\tau)$
evaluated at correlation time between $5<\tau/\,\mrm{ps}<10$.

\subsection{System size}

\begin{figure}[H]
    \centering
    \includegraphics[width=\linewidth]{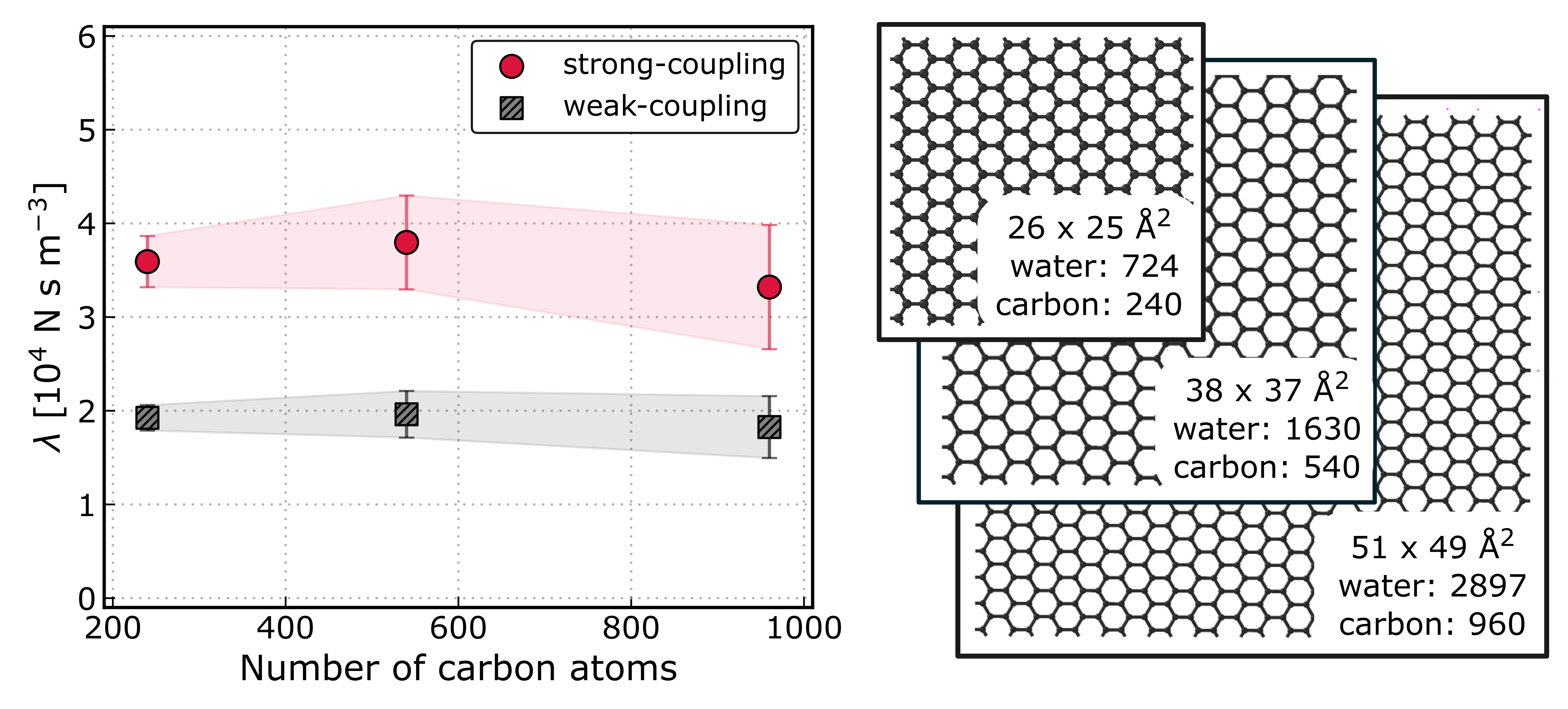}
    \caption{\textbf{Sensitivity of friction coefficient to system size.}
    Statistical errors are obtained from block-averaging.
    The three system sizes tested are illustrated on the 
    right panel where only the solid sheets are shown.
    The corresponding total area for the sheet and the
    number of water molecules and carbon atoms are given 
    for each system}
    \label{si:friction_size}
\end{figure}

To make sure the employed system size is sufficient to obtained 
converged values, we checked if our results remain consistent at 
larger system sizes. 
In Fig.~\ref{si:friction_size}, the friction coefficient for 
two representative Drude masses are shown for systems with
different lateral areas in the $(x,y)$ plane. 
The additional simulations were carried out on an interface
with approximately the same liquid film thickness 
$\approx\SI{37}{\angstrom}$. 
The friction coefficients computed from these simulations
are almost identical to those presented in the main article.

\subsection{Simulation time}

\begin{figure}[H]
    \centering
    \includegraphics[width=0.55\linewidth]{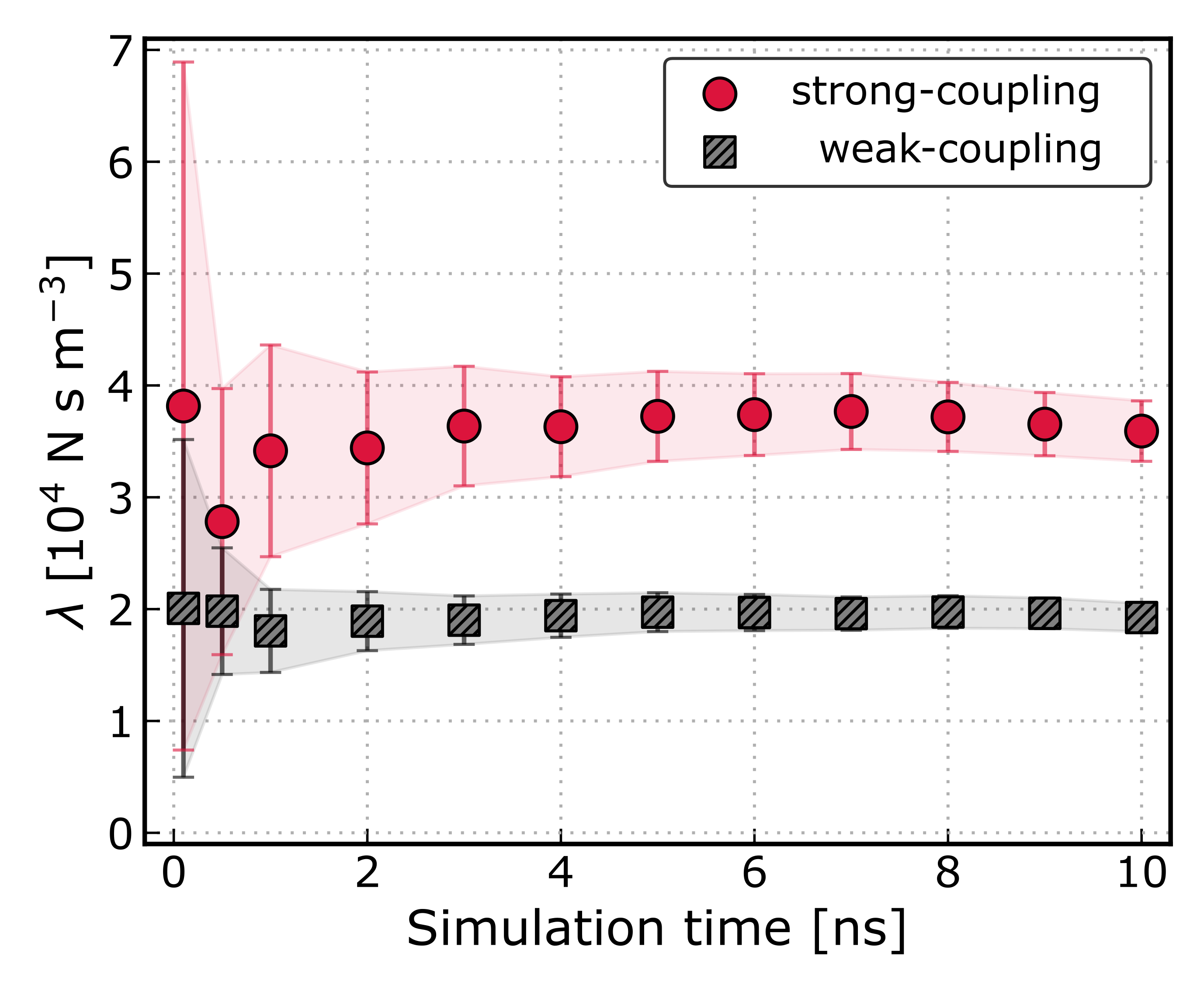}
    \caption{\textbf{Sensitivity of friction coefficient to simulation time.}
    $\lambda$ changes relatively little with increasing simulation time.
     Statistical errors are obtained from block-averaging.}
      \label{si:friction_simulationtime}
\end{figure}

Analogous to checking the impact of system size, we
checked the convergence of the friction coefficient 
with the simulation time length, as presented in 
Fig.~\ref{si:friction_simulationtime}.
For all cases, the friction coefficient is found to 
change relatively little with increasing
simulation time. 
While statistical errors are larger for shorter simulations,
simulation times as short as \SI{1}{\nano\second} is enough 
to converge friction to within 10\%.
Therefore, the employed simulation time of \SI{10}{\nano\second}
is sufficient for a converged friction coefficient.

\subsection{Time-step}

\begin{figure}[H]
    \centering
    \includegraphics[width=\linewidth]{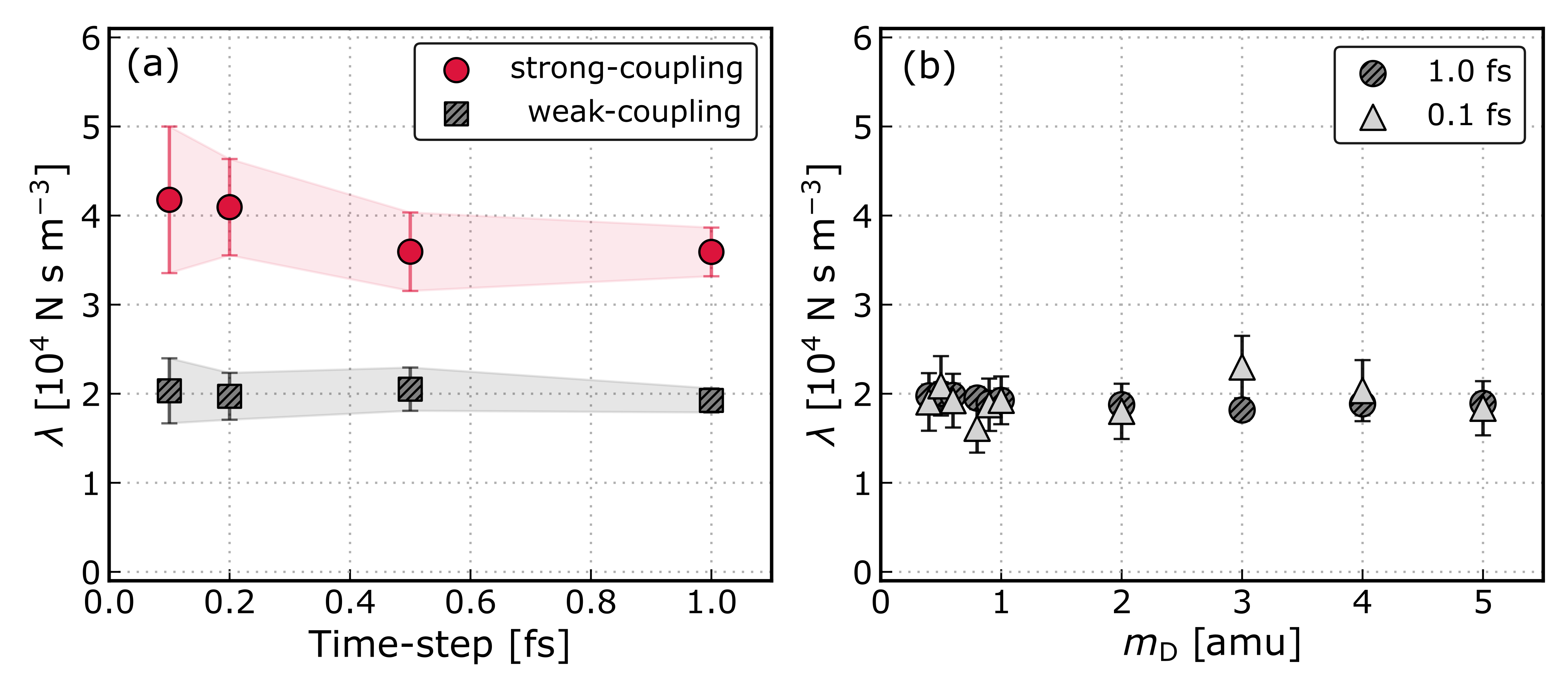}
    \caption{\textbf{Sensitivity of friction coefficient to time-step.}
    (a) Dependence of the computed $\lambda$ on the time-step used in 
    simulations.
    (b) Comparison of $\lambda$ for simulations with 
    $m_{\mrm{D}}\lesssim  5\,\mrm{amu}$ using
    a time-step of $0.1\,\mrm{and}\,1\,\mrm{fs}$.
    Statistical errors are obtained from block-averaging.
    }
    \label{si:friction_timestep}
\end{figure}

We test the sensitivity of the friction coefficients in both the 
weak-coupling and strong-coupling regimes with the time-step 
used in simulations and see that the results agree well for 
time-step of $0.1, 0.2, 0.5\,\mrm{and}\,1\,\mrm{fs}$, as shown 
in Fig.~\ref{si:friction_timestep}(a). 
However, when dealing with small Drude masses, problems with 
energy drifts often arise due to the inherently high frequency
of the individual Drude oscillators\cite{Lamoureux2003}. 
We therefore also check that the time-step of
$\SI{1}{\femto\second}$ employed gives acceptable value of 
$\lambda$ for Drude masses of $0.4 \leq m_{\mrm{D}}/\mrm{amu} \leq  5$
in our simulations, as shown in Fig.~\ref{si:friction_timestep}(b).
Here, all these simulations belong to the weak-coupling regimes.

\begin{figure}[H]
    \centering
    \includegraphics[width=\linewidth]{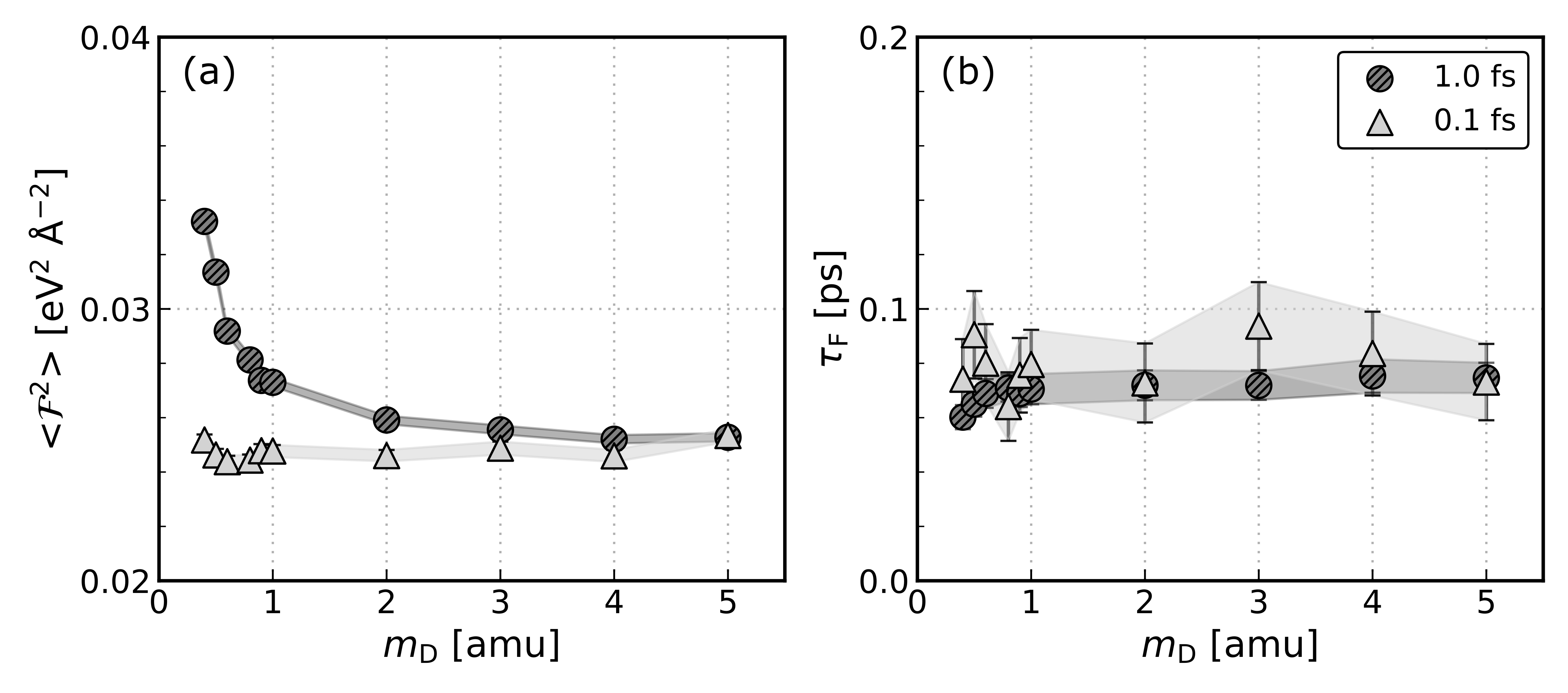}
    \caption{\textbf{Sensitivity of the static and dynamical
    components of friction to time-step.} 
    Variation of (a) the mean squared force $\langle \mcl{F}^2 \rangle$ 
    and (b) the force decorrelation time $\tau_{\mrm{F}}$ for 
    simulations with Drude masses $0.4 \leq m_{\mrm{D}}/\mrm{amu} \leq  5$
    using a time-step of $0.1\,\mrm{and}\,1\,\mrm{fs}$.}
    \label{si:friction_decompose_timestep}
\end{figure}

Further analysis by extracting the static component, quantified by 
$\langle \mcl{F}^2 \rangle$, and the dynamical component,
quantified by $\tau_{\mrm{F}}$, of the friction is shown in
Fig.~\ref{si:friction_decompose_timestep}.
This reveals that although $\lambda$ converge
for $m_{\mrm{D}}\lesssim  2\,\mrm{amu}$ for a time-step 
of $\SI{1}{\femto\second}$, $\langle \mcl{F}^2 \rangle$
and $\tau_{\mrm{F}}$ diverge for these small masses.
Therefore, in the main article, $\langle \mcl{F}^2 \rangle$
and $\tau_{\mrm{F}}$ for $m_{\mrm{D}}\lesssim  2\,\mrm{amu}$
cases are computed from simulations using a time-step of
$\SI{0.1}{\femto\second}$.

\subsection{Thermostats}

\begin{figure}[H]
    \centering
    \includegraphics[width=\linewidth]{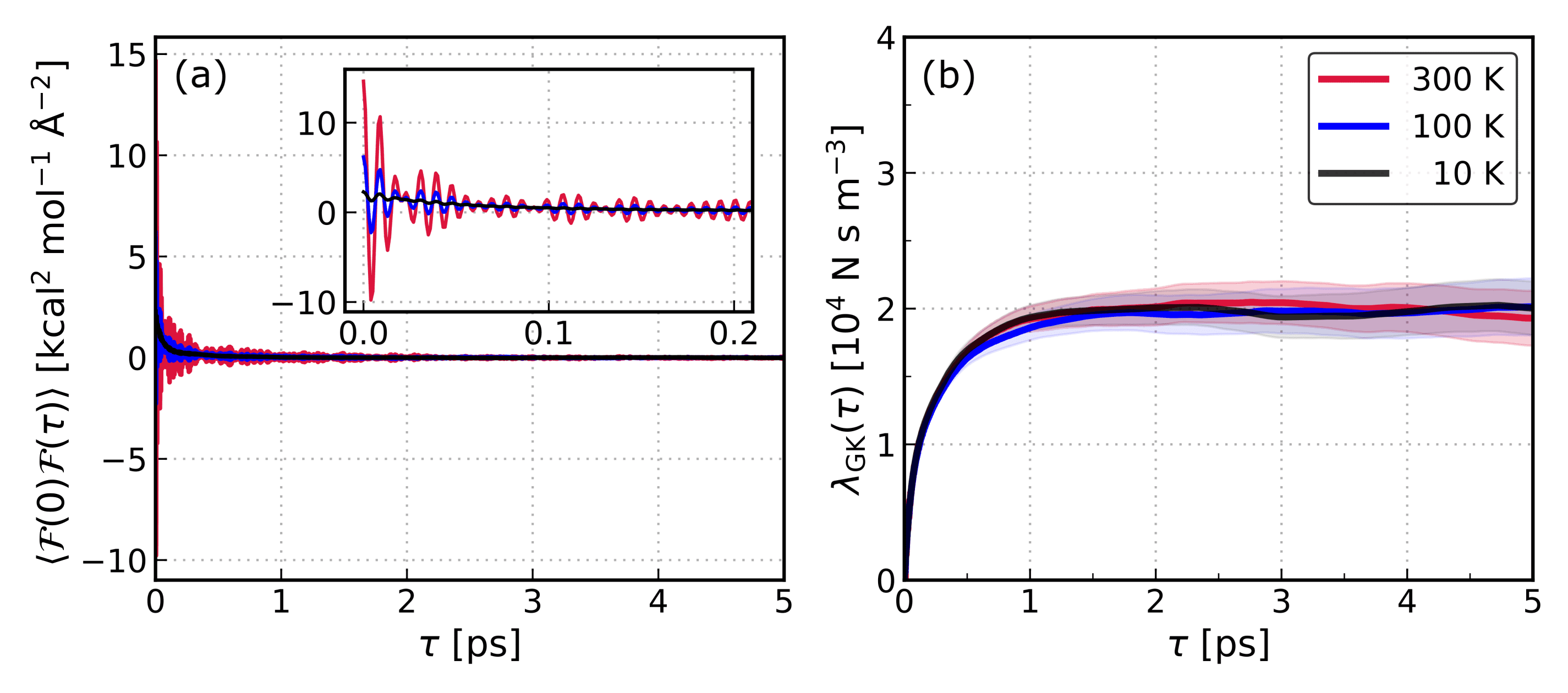}
    \caption{\textbf{Senstivity of friction coefficient to the
    temperature of the Drude particles.}
    (a) The force autocorrelation function for simulations
    with $m_{\mrm{D}}=1\,\mrm{amu}$ where the Drude particles
    are kept at $T_{\mrm{sol}} = 10, 100, 300 \,\mrm{K}$.
    The inset shows clearer the oscillations at short times.
    (b) The convergence of $\lambda_{\mrm{GK}}$ for the three cases 
    where statistical errors (shaded area) are obtained from block-averaging.
    } 
    \label{si:friction_Tdrude}
\end{figure}

Ordinarily, in simulations of polarizable systems using the classical Drude 
oscillator model, the temperature of the Drude particles is often 
kept low \cite{Lamoureux2003,Schroder2010,Rupakheti2020}
to minimize energy exchange between the nuclear and Drude motion.
In this work, we found numerical instabilities in simulations with 
low $T_{\mrm{sol}}$ for $m_{\mrm{D}} \gtrsim 10\,\mrm{amu}$.
Therefore, for all results presented in this work, we used two separate 
thermostats to keep the temperatures of the water, $T_{\mrm{wat}}$,
and the Drude particles, $T_{\mrm{sol}}$, both at $\SI{300}{\kelvin}$.
We carried out simulations with $m_{\mrm{D}}=1\,\mrm{amu}$ 
(the weak-coupling regime) using different values for
$T_{\mrm{sol}} = 10, 100\,\mrm{and}\,300 \,\mrm{K}$.
As shown in Fig.~\ref{si:friction_Tdrude}, while the force 
autocorrelation function shows oscillations of greater magnitude 
at higher $T_{\mrm{sol}}$, its integral and therefore the
friction coefficient is not significantly affected.

\begin{figure}[H]
    \centering
    \includegraphics[width=\linewidth]{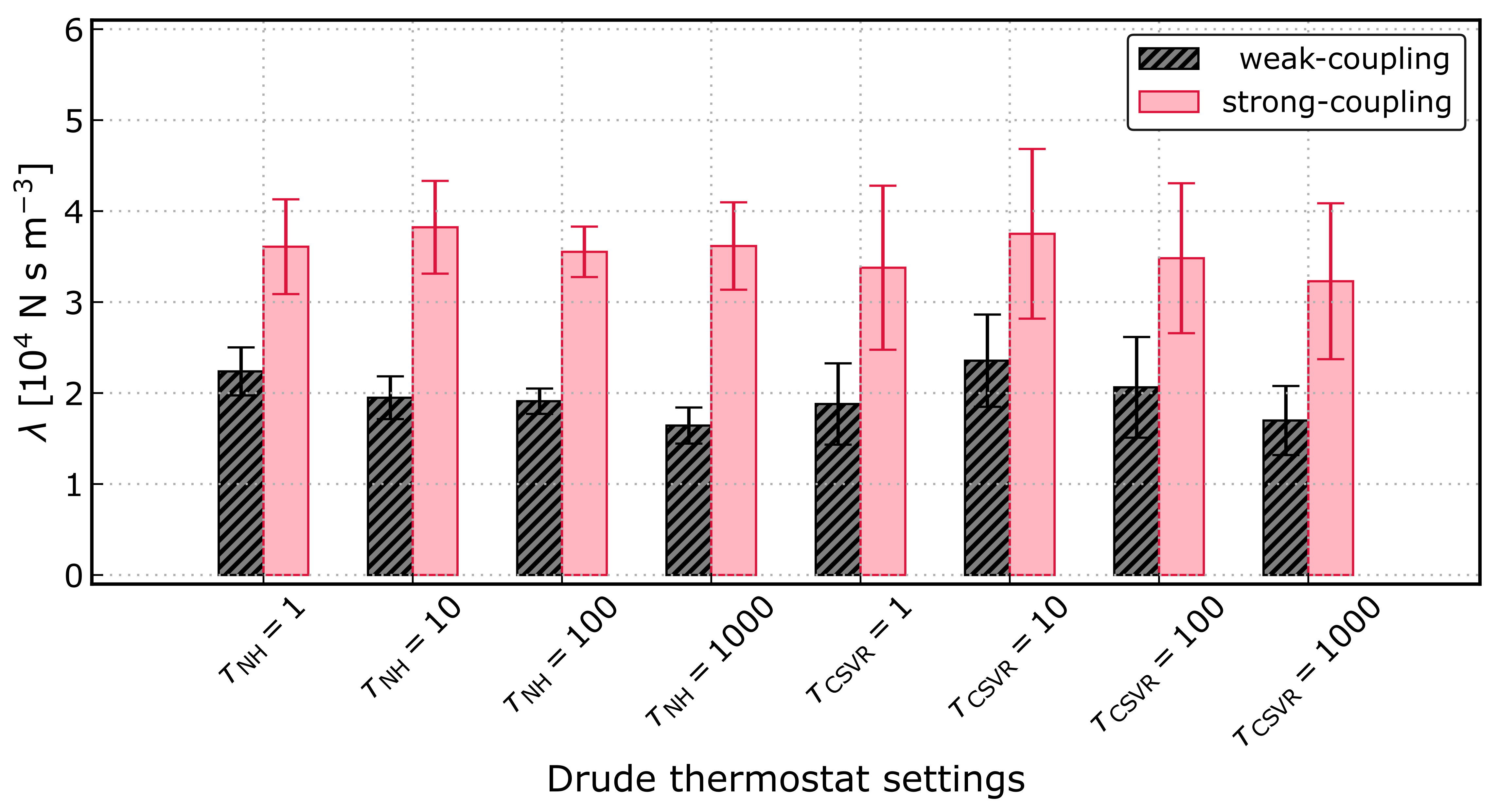}
    \caption{\textbf{Sensitivity of friction coefficient to thermostat 
    settings for Drude particles.}
    The values of the damping time constant of the thermostat 
    ($\tau_{\mrm{NH}}$ for the Nos\'{e}--Hoover thermostat and
    $\tau_{\mrm{CSVR}}$ for the CSVR thermostat)
    for the Drude particles are given in fs.
    Statistical errors are obtained from block-averaging.}
    \label{si:friction_thermostat}
\end{figure}

We also test the sensitivity of the friction coefficient to 
the thermostat settings used to maintain the temperature of the 
Drude particles.
In Fig.~\ref{si:friction_thermostat}, we show that the friction
coefficients agree well for different damping times of the
Nos\'{e}--Hoover thermostat for the Drude particles
$\tau_{\mrm{NH}} = 1, 10, 100, 1000 \,\mrm{fs}$
in both the weak-coupling and strong-coupling regimes.
Equivalent results were also obtained with the canonical 
sampling through velocity rescaling (CSVR) thermostat\cite{Bussi2007}.

\subsection{Liquid film thickness}

\begin{figure}[H]
    \centering
    \includegraphics[width=\linewidth]{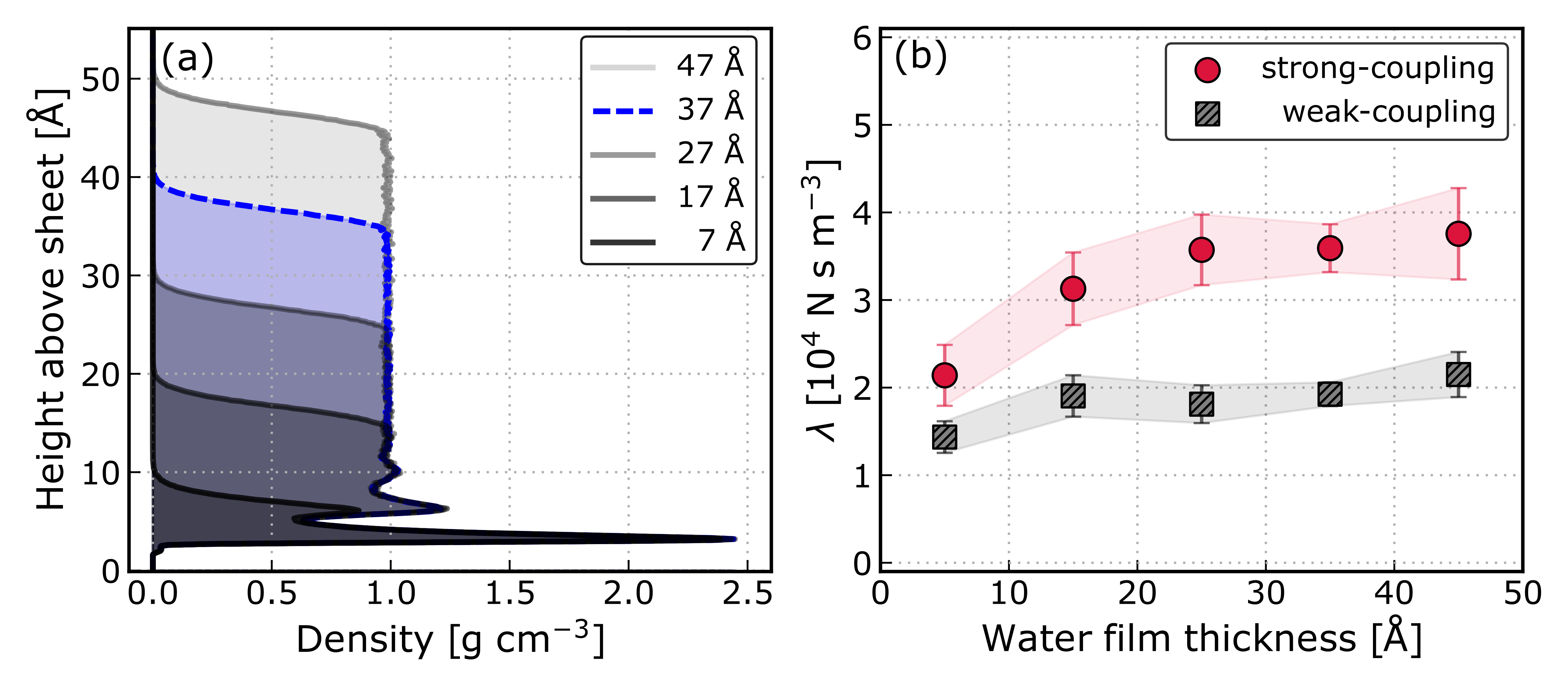}
    \caption{\textbf{Sensitivity of friction to simulation 
    time to the liquid film thickness.}
    (a) The planar density profile of different film 
    thicknesses above the flat solid sheet is shown for 
    different water film thicknesses (indicated in the legend). 
    In the main article, we use a thickness of $\approx 37 \, \angstrom$,
    as indicated by the dashed blue line.
    (b) The sensitivity of friction to the liquid thickness. 
    Statistical errors are obtained from block-averaging.}
    \label{si:friction_thickness}
\end{figure}

Simulations for systems with varying thickness of the 
water film were performed to test the convergence of 
the friction coefficient.
In Fig.~\ref{si:friction_thickness}(a), we show the planar
average mass density profiles for the liquid--solid
system with a water film of thickness 
$\approx 7, 17, 27, 37$ and $47 \, \angstrom$.
The thickness is determined by the height from the
positions of the carbon atoms in the sheet to where
the water density at the liquid--vapour interface is 
equal to $\SI{0.5}{\gram\per\centi\meter\cubed}$.
In Fig.~\ref{si:friction_thickness}(b), we show 
how the extracted friction coefficient changes with the 
water film thickness.
A water film with thickness $\gtrsim 17 \, \angstrom$ is 
required to have a region with bulk mass density
and a converged friction coefficient.
Therefore, employing a thickness of $\approx 37 \, \angstrom$
for simulations of our main results is justified.

\subsection{Electrostatic boundary conditions}

\begin{figure}[H]
    \centering
    \includegraphics[width=0.55\linewidth]{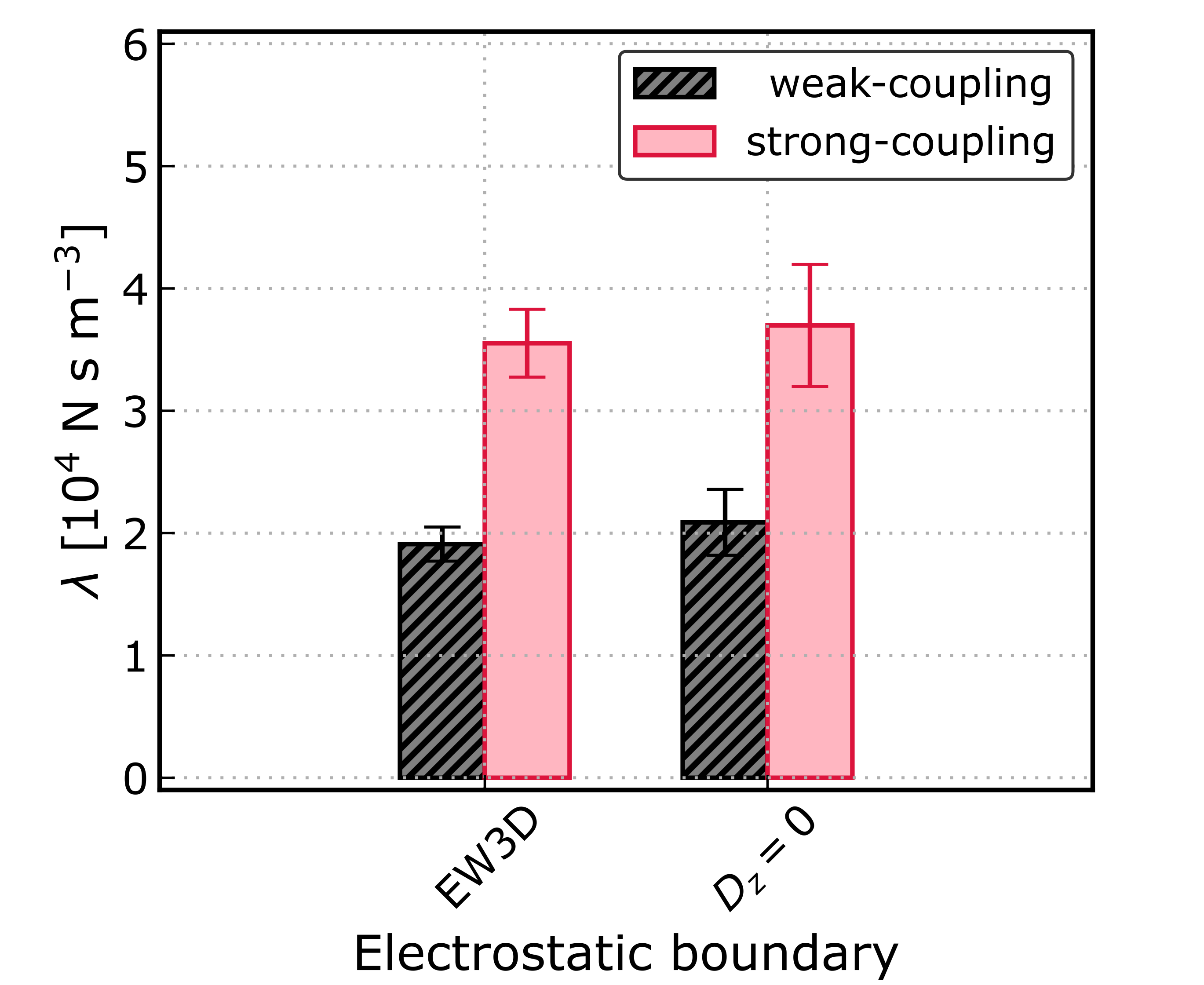}
    \caption{\textbf{Sensitivity of friction coefficient to electrostatic 
    boundary conditions.}
    The friction coefficients in the weak-coupling and strong-coupling
    agree well for simulations employing the EW3D method 
    and the $D_z=0$ method. Statistical errors are obtained from 
    block-averaging.}
     \label{si:friction_ewald}
\end{figure}

For simulations presented in the main article,
we treated electrostatics by applying the conventional 
three-dimensional Ewald summation (EW3D) technique\cite{Shelly1996}. 
This method is commonly employed in simulations of different
interfacial systems \cite{Alejandre1995,Feller1996,Poggioli2021}.
To test the influence of this choice to the friction coefficient,
we performed additional simulations with hybrid boundary conditions
in which the Ewald summation is applied to the $x$ and $y$ directions
while the electric displacement field in the $z$ direction is set to 
zero ($D_z=0$).
This is done using the finite field approach \cite{Zhang2016,Zhang2017,Cox2019}
and such hybrid boundary conditions have been shown 
\cite{Zhang2016} to be formally equivalent to the 
Yeh--Berkowitz correction \cite{Yeh-Berkowitz1999} 
that decouples the electrostatic interactions between a slab of 
material and its periodic images.
We found good agreement for the friction coefficients of both
the weak-coupling and strong-coupling regimes between the EW3D
and the $D_z=0$ methods, as shown in Fig.~\ref{si:friction_ewald}.

\newpage

\section{Sensitivity of the dependence of friction on solid charge density frequency}

For the simulations presented in the main article, we fixed 
$Q_{\rm D}\,=\,1.852\,e$ and $k_{\mrm{D}}\,=\,1000\,\mrm{kcal}\,\mrm{mol}^{-1}\angstrom^{-2}$
for the Drude oscillators, which have been parameterized by
Misra and Blanckstein to recover the polarizability tensor of a 
periodic graphene lattice \cite{Misra2017}. 
In principle, $Q_{\mrm{D}}$ and $k_{\mrm{D}}$ both control
the solid charge density and changing their values will affect
the coupling between the solid and the liquid charge densities,
and therefore the friction.
Here we will present a check for the sensitivity of the dependence 
of friction on solid charge frequency upon changing to different
values for each of these parameters.

\subsection[Varying the Drude charge]{Varying the Drude charge}

\begin{figure}[H]
    \centering
    \includegraphics[width=0.65\linewidth]{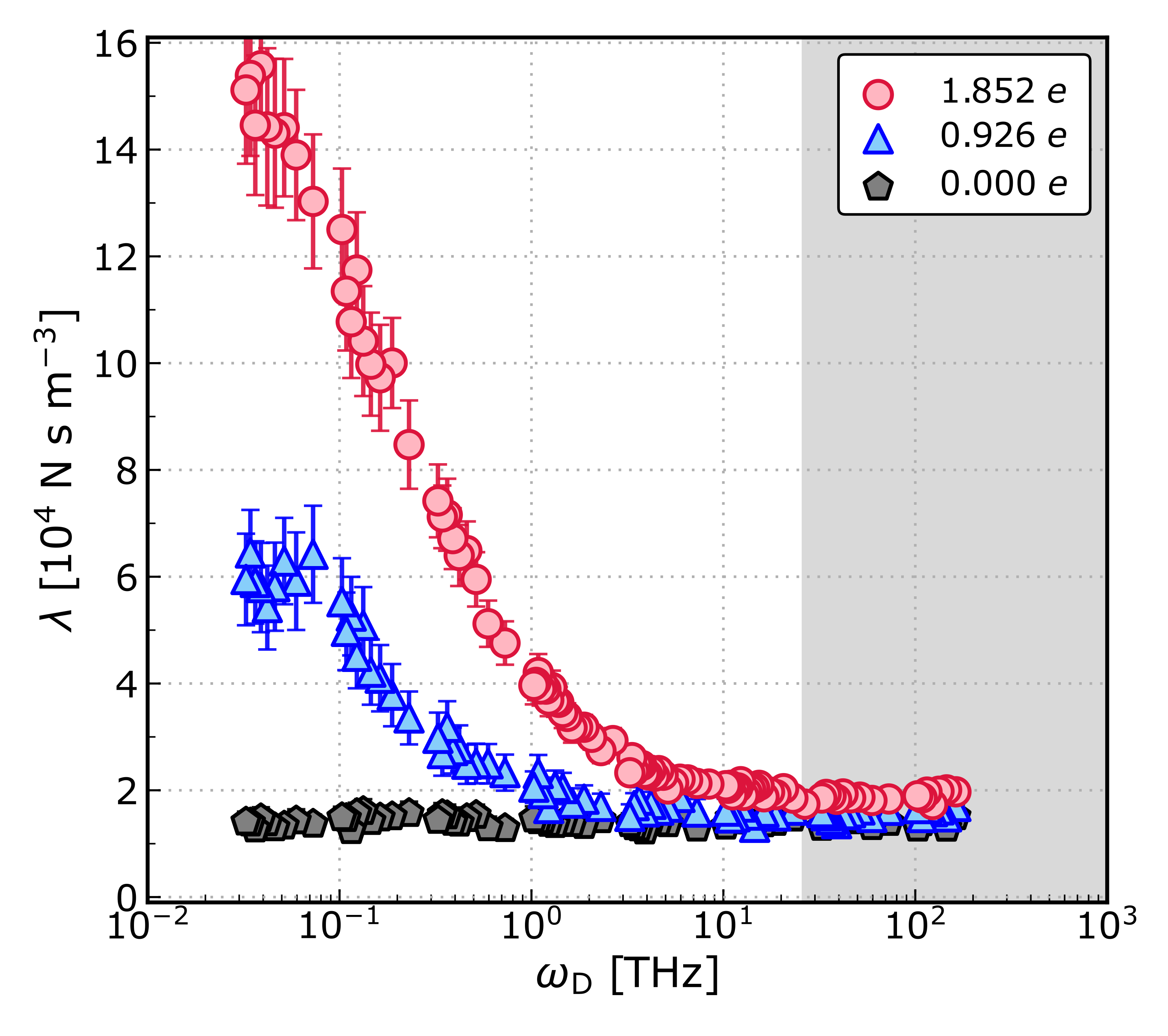}
    \caption{\textbf{Dependence of friction on solid charge 
    density frequency for different Drude charges.} 
    Values of $Q_\mrm{D}$ are indicated in the legend and $k_{\mrm{D}}\,=\,1000\,\mrm{kcal}\,\mrm{mol}^{-1}\angstrom^{-2}$ in all cases.
    The weak-coupling and strong-coupling regimes are shaded grey and not shaded, respectively.
    For simulations with $Q_\mrm{D}=0$,
    $\lambda$ remains constant with $\omega_{\mrm{D}}$.
    For simulations with $Q_\mrm{D}=0.926$ and $1.852\,e$, $\lambda$
    increases as $\omega_{\mrm{D}}$ is decreased in the strong-coupling regime.}
    \label{si:friction_qD}
\end{figure}

We performed two additional sets of simulations of the liquid--solid interface
with $Q_\mrm{D}=\,0$ and $0.926\,e$ while keeping 
$k_{\mrm{D}}\,=\,1000\,\mrm{kcal}\,\mrm{mol}^{-1}\angstrom^{-2}$ 
for Drude masses in the range
$1\lesssim m_{\rm D}/{\rm amu} \lesssim 10^7$ while other aspects of the
simulations are kept the same.
In the main article, we show the dependence of the friction on
the frequency of the solid charge density by plotting $\lambda$
against $\omega_{0}$.
Based on the discussion in Sec.~\ref{sec:coupling}, 
we can approximate the solid charge density as
$\omega_{0} \approx \omega_{\mrm{D}}$ instead.
Therefore, for convenience, in these additional analyses,
$\lambda$ is plotted against $\omega_{\mrm{D}}$ as shown in Fig.~\ref{si:friction_qD}.
The relationship mapped out for $Q_\mrm{D}=\,1.852\,e$ is essentially unchanged 
compared to the one presented in Fig.~2(a) in the main article.

For $Q_\mrm{D}=\,0$, there is no charge density in the solid, i.e 
$\tilde{n}_\mrm{sol}(q,\omega)=0$, so the only contribution
to friction is from the surface roughness. 
Therefore the friction remains constant with $\omega_{\mrm{D}}$
at $\lambda\approx1.7\times10^4\,\si{\Nsm}$
Increasing $Q_\mrm{D}$ to $0.926$ and $1.852\,e$ slightly
increases the surface roughness contribution to friction,
as seen from the flattening in the weak-coupling regime. 
Meanwhile, the contribution to friction from charge density
coupling increases much more significantly in the 
strong-coupling regime for higher values of $Q_\mrm{D}$.
This behavior supports the fact that the increase in friction
at the low frequency end is indeed due to coupling of charge
density between the solid and the liquid and larger charge density
in the solid will couple more strongly with the liquid,
resulting a larger increase in friction the strong-coupling regime.

\subsection{Varying the spring constant}

Here, we performed two additional sets of simulations of the 
liquid--solid interface with $k_\mrm{D}=\,600$ and $1000\,\mrm{kcal}\,\mrm{mol}^{-1}\angstrom^{-2}$
while keeping $Q_\mrm{D}=1.852\,e$,
also for Drude masses in the range $1\lesssim m_{\rm D}/{\rm amu} \lesssim 10^7$,
while other aspects of the simulations are kept the same.

\begin{figure}[H]
    \centering
    \includegraphics[width=0.65\linewidth]{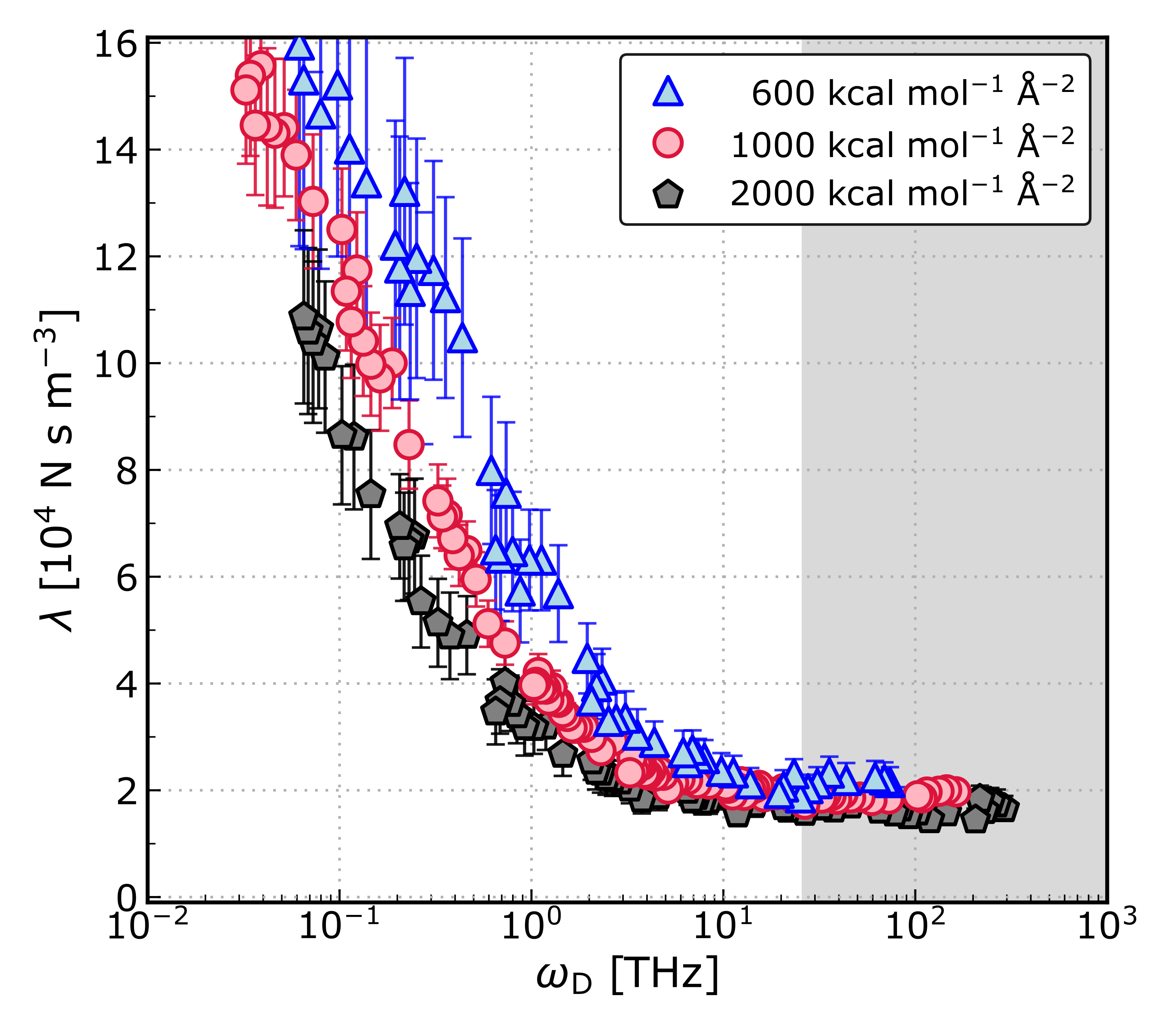}
    \caption{\textbf{Dependence of friction on solid charge 
    density frequency for different harmonic force constant.}
    Values of $k_\mrm{D}$ are indicated in the legend and
    $Q_\mrm{D}=\,1.852\,e$ in all cases.
    The weak-coupling and strong-coupling regimes are shaded grey and not shaded, respectively.
    For all cases, $\lambda$
    increases once $\omega_{\mrm{D}}\lesssim\omega_{\mrm{lib}}$ 
    in the strong-coupling regime.}
    \label{si:friction_kD}
\end{figure}

Again, since the solid charge density is changed when
$k_\mrm{D}$ is varied, differences in the absolute values
of $\lambda$ are expected and indeed observed, as shown
in Fig.~\ref{si:friction_kD}.
According to Eq.~\ref{eq:polarizability}, a decrease in $k_\mrm{D}$ 
means that the atom modeled by the Drude oscillator becomes less 
polarizable. 
Therefore, the friction contribution
due to the surface roughness is higher for lower values of $k_\mrm{D}$ 
but the differences in the $\lambda$ in the weak-coupling regime
are relatively small.
The contribution to friction from charge density coupling 
is observed to be more strongly affected since $\lambda$
increases much more sharply for lower $k_{\mrm{D}}$
as $\omega_{\mrm{D}}$ is decreased in the strong-coupling regime.
The important thing to stress is that this increase of $\lambda$ 
due to charge density coupling in all cases occur 
once $\omega_{\mrm{D}}\lesssim\omega_{\mrm{lib}}$,
supporting the separation into the weak-coupling and strong-coupling
regimes in the main article.

\subsection{Flexible water model}

\begin{figure}[H]
    \centering \includegraphics[width=\linewidth]{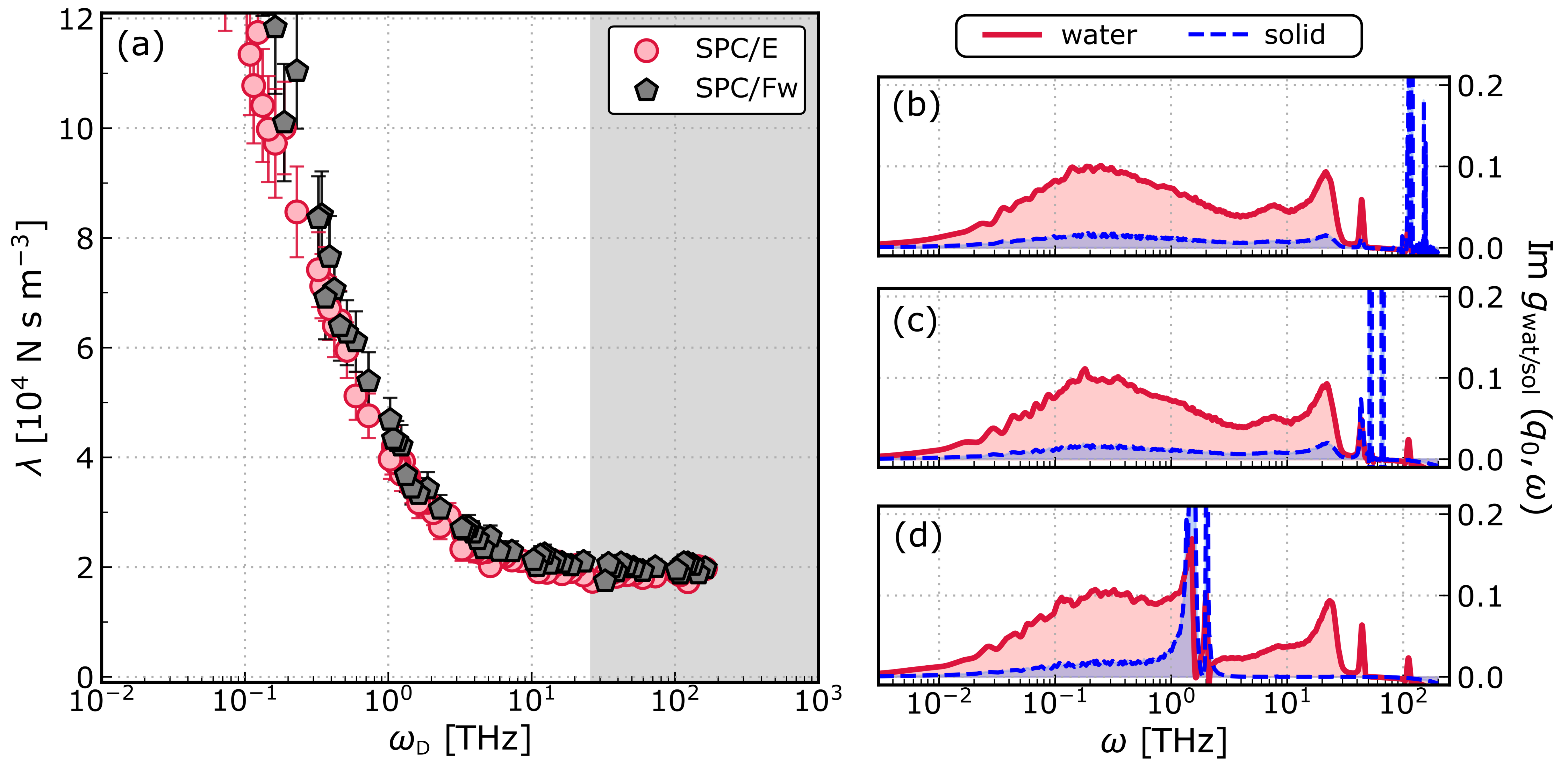} 
    \caption{\textbf{Dependence
    of friction on solid charge density frequency for a flexible water
    model.} (a) The friction dependence on $\omega_{\rm D}$ remains
    almost unchanged when simulations are performed with a rigid
    (SPC/E) or flexible (SPC/Fw) water model. When solid principal
    peaks coincide with (b) the OH stretching modes or (c) the
    in-plane stretching mode, there is no significant response in
    either the solid or the liquid. (d) The strong-coupling regime
    remains at frequencies lower than the librational
    peak.}  \label{si:friction_flexiblewat}
\end{figure}

To check the sensitivity of our results to the presence of
intramolecular modes in water, we performed additional simulations
with a flexible water model SPC/Fw\cite{Wu2006}.
In additional to the intermolecular librational and Debye modes,
SPC/Fw also captures the OH stretching modes as a peak centered at
$\approx\SI{100}{\THz}$ and in-plane bending modes as a peak centered
at $\approx\SI{50}{\THz}$.
The addition of these peaks in the water dielectric spectrum does not
affect the conclusions drawn in the main article, as shown in Fig.~\ref{si:friction_flexiblewat}.

\subsection{Phonon contribution}

\begin{figure}[H]
    \centering
    \includegraphics[width=\linewidth]{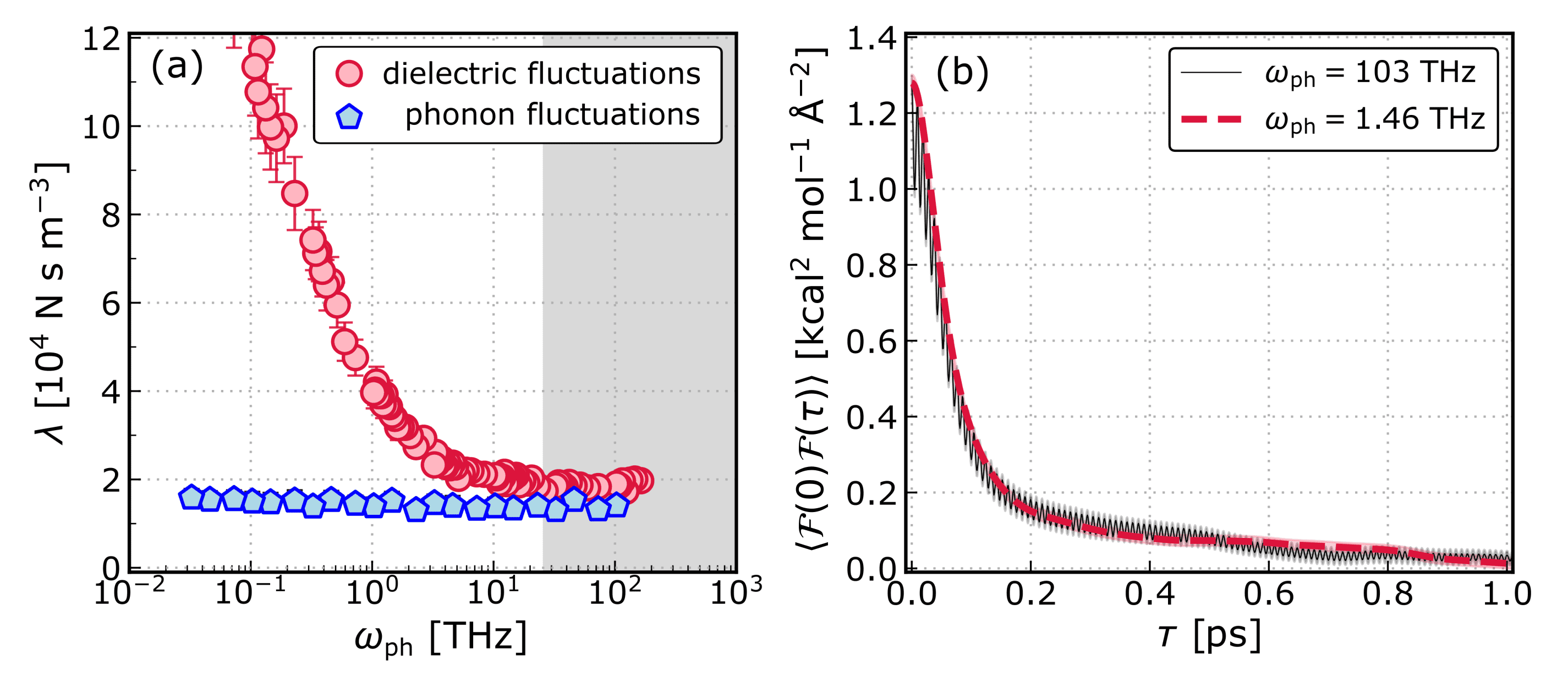}
    \caption{\textbf{Dependence of the friction on the phonon
    mode frequency.} (a) The friction is not strongly affected by
    phonon fluctuations compared to the effect seen for dielectric 
    fluctuations in the the solid. (b) The force--force autocorrelation
    functions are shown for two different phonon frequencies as indicated in the legend. Unlike for dielectric modes, the phonon modes do not change the 
    force decorrelation time significantly.}
    \label{si:friction_lj}
\end{figure}

In this work, we do not consider the effect of phonons in the solid on
the friction of the interface, which would in principle affect both
the surface roughness contribution $\lambda_{\mrm{SR}}$ and the
contribution from the coupling of the dynamics of the solid and the
liquid $\lambda_{\mrm{THz}}$. Here, we can simply investigate how a
single phonon mode in the solid would change the friction. We do this
by considering a simple model in which the solid is a set of
independent harmonic oscillators whereby the carbon atoms are now
attached to their lattice positions via a harmonic spring with force
constant $k_{\mrm{ph}}$, and ascribed a mass $m_{\rm ph}$. The carbon
atoms still interact with the water through the same Lennard-Jones
potential. By varying $m_{\rm ph}$ we can tune the phonon frequency
$\omega_{\mrm{ph}}=(k_{\mrm{ph}}/m_{\mrm{ph}})^{1/2}$ (with
$k_{\mrm{ph}}=\SI{1000}{\kilo\cal\per\mol\per\angstrom\squared}$) in a
similar fashion to how we tuned the solid's dielectric spectrum. In
contrast to varying $m_{\rm D}$, we see that friction is relatively
insensitive to changes in $m_{\rm ph}$, as seen in Fig.~\ref{si:friction_lj}.
This result is in line with the prediction of QF theory that, for
water at carbon substrates, phonon contributions to quantum friction are relatively small.

\newpage
\section{Coupling of charge densities}
\label{sec:coupling}

In this section, we present further analyses of the
surface response functions of the solid and the liquid in
support of the conclusions made on the separation
into the weak-coupling and strong-coupling regimes
in the main article.

\subsection{Solid surface response function in the absence of water}

In the main article, we show that the surface response function of the solid 
$g_{\mrm{sol}}(q_0,\omega)$ is dominated by two peaks that are slightly blue-shifted from
$\omega_{\rm D}$.
Here we provide more details into the origin of each of these peaks, with
the peak lower in frequency denoted as $\omega_0$ and the peak higher
in frequency denoted as $\omega_1$.

\begin{figure}[H]
    \centering
    \includegraphics[width=0.55\linewidth]{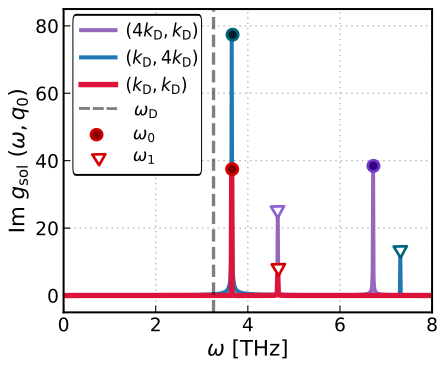}
    \caption{\textbf{Solid surface charge density in the absence of water.}
    The surface response function $g_{\mrm{sol}}(q_0,\omega)$
    of a carbon sheet with Drude particles of mass $m_{\mrm{D}}=10^3\,\mrm{amu}$
    such that $\omega_{\mrm{D}}=(k_{\mrm{D}}/m_{\mrm{D}})^{1/2}=3.3\,\mrm{THz}$
    in the absence of water.
    The value of $\omega_{\mrm{D}}$ is marked with a dashed vertical line.
    When the spring is changed from having an isotropic force constant
    $(k_{{xy}}, k_{{z}})=(k_{\mrm{D}}, k_{\mrm{D}})$
    to an anisotropic one, the modes in $g_{\mrm{sol}}(q_0,\omega)$ are shifted
    accordingly. When $(k_{{xy}}, k_{{z}})=(4k_{\mrm{D}}, k_{\mrm{D}})$,
    the parallel mode at $\omega_0$ (denoted with a filled circle) is shifted.
    When $(k_{{xy}}, k_{{z}})=(k_{\mrm{D}}, 4k_{\mrm{D}})$,
    the perpendicular mode at $\omega_1$ (denoted with an empty triangle) is shifted.
    }
    \label{si:solid_chargedensity}
\end{figure}

In the absence of the Coulomb interaction, the Drude oscillators are
simply a set of independent harmonic oscillators characterized by a
single frequency $\omega_{\rm D}$. In the presence of the Coulomb
interaction, the motion of the Drude oscillators is no longer
isotropic, with motion perpendicular to the plane of the graphene
sheets having a higher frequency ($\omega_1$) than parallel motion
($\omega_0$). To illustrate this point further, in
Fig.~\ref{si:solid_chargedensity}(b), we show how
$g_{\mrm{sol}}(q_0,\omega)$ changes as the force constants of the
springs are changed to $(k_{{xy}}, k_{{z}})= (4k_{\mrm{D}},
k_{\mrm{D}})$ and $(k_{{xy}}, k_{{z}}) = (k_{\mrm{D}},
4k_{\mrm{D}})$, where
$k_{\mrm{D}}=\SI{1000}{\kilo\cal\per\mol\per\angstrom\squared}$. (To
be clear, in the main article, an isotropic spring constant $k_{\rm
D}$ is used throughout.)
When $(k_{xy}, k_{z})= (4k_{\mrm{D}}, k_{\mrm{D}})$,
$\omega_0$ shifts to higher frequency and $\omega_1$ is unaffected.
Conversely, when $(k_{{xy}}, k_{{z}}) = (k_{\mrm{D}},
4k_{\mrm{D}})$, $\omega_1$ shifts to higher frequency and $\omega_0$
position is unaffected.

%
%
%
%
%
%

\begin{figure}[H]
    \centering
    \includegraphics[width=\linewidth]{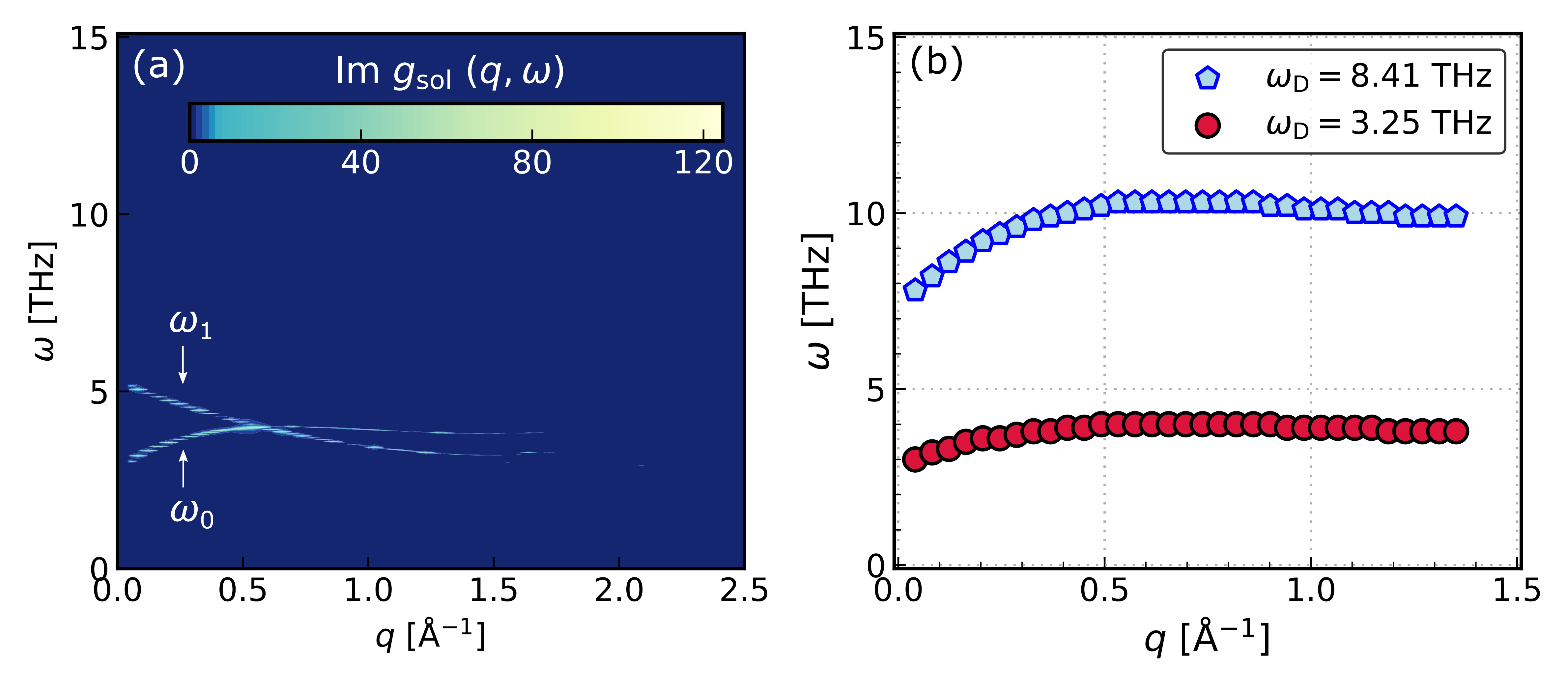}
    \caption{\textbf{Dispersion relation of the solid modes.}
    (a) The solid surface response $g_{\mrm{sol}}(q,\omega)$ of a carbon sheet with Drude particles of mass $m_{\mrm{D}}=10^3\,\mrm{amu}$ 
    in the absence of water.
    The tangential mode increases in frequency while the perpendicular mode
    decreases in frequency before flattening out up to $q_{\mrm{max}}\approx1.5\angstrom^{-1}$.
    (b) The solid tangential mode shows relatively flat dispersion
    relation, as shown for two representative simulations with 
    $\omega_{\mrm{D}}$ indicated in the legend.}
    \label{si:width_dispersion}
\end{figure}

To access the dispersion relation of the solid modes in more detail,
we perform simulations of the solid system with a supercell
length of $\SI{153.36}{\angstrom}$. 
In Fig.~\ref{si:width_dispersion}(a), we show $g_{\mrm{sol}}(q,\omega)$ in
both $q$-space and $\omega$-space.
Both the solid modes show relatively flat dispersion, with very narrow widths 
in $\omega$-space and spanning up to $q_{\mrm{max}}\approx1.5\angstrom^{-1}$ in $q$-space.
This flat dispersion of the solid modes is seen in simulations with different
values of $\omega_{\mrm{D}}$, as illustrated in
Fig.~\ref{si:width_dispersion}(b).

\subsection{Solid surface response function in the presence of water}

In the absence of water, there are two sources of dissipation in the solid
as the Drude oscillators are coupled to a thermostat and also
interact with each other by electrostatic interactions. 
Therefore, the principal modes in the solid have finite widths in the
response function. 
However, since the Drude oscillators are only weakly-coupled, they
remain underdamped and these widths remain very small. 
In the presence of water, when there is strong-coupling with the
intermolecular modes of water, the Drude oscillators are more
strongly damped, leading to the broadening of their widths.

Focusing on the case when $\omega_{\mrm{D}}=\SI{3.26}{\THz}$, we 
can quantify this broadening by fitting the surface response
function $g_{\mrm{sol}}(q_0,\omega)$ obtained from simulations to
a double-Lorentzian
\begin{equation}
    \mrm{Im}\,g_{\mrm{sol}}(q_0,\omega) = \frac{a_0 \eta_0^2 }{(\omega-\omega_0)^2 + \eta_0^2} 
    + \frac{a_1 \eta_1^2 }{(\omega-\omega_1)^2 + \eta_1^2},
\label{eq:lorentzian}
\end{equation}
where $a_0$ is the amplitude, $\omega_0$ is the centered frequency, $\eta_0$
is the width of the tangential mode and $a_1$, $\omega_1$, $\eta_1$ are similarly
defined for the perpendicular mode.
The results obtained for the fitting parameter are given in
Table~\ref{Lorentzianfit}. We see that both of the principal
modes experience a small redshift to lower frequency of 
$\approx\SI{0.06}{\THz}$. More prominently, the amplitudes of
both peaks are reduced and the widths are broadened in the presence
of water.

\begin{table}[H]
    \centering
    \begin{tabular}{>{\color{black}}c >{\color{black}}c >{\color{black}}c >{\color{black}}c >{\color{black}}c >{\color{black}}c >{\color{black}}c}
    \hline
    \hline
     &  $a_0$  & $\omega_0$ [THz] & $\eta_0$ [THz] & $a_1$ & $\omega_1$ [THz]  & $\eta_1$ [THz]\\
    \hline
     without water &  17  & 4.65  & 0.0032 & 36 & 3.65 & 0.0020 \\
     with water &  0.68   & 4.59  & 0.154 & 0.82 & 3.49 & 0.1742 \\
    \hline
    \hline
\end{tabular}
\caption{\tbf{Parameters for fitting the solid surface response to a double-Lorentzian as Eq.~\ref{eq:lorentzian}.}}
\label{Lorentzianfit}
\end{table}

\begin{figure}[H]
    \centering
    \includegraphics[width=\linewidth]{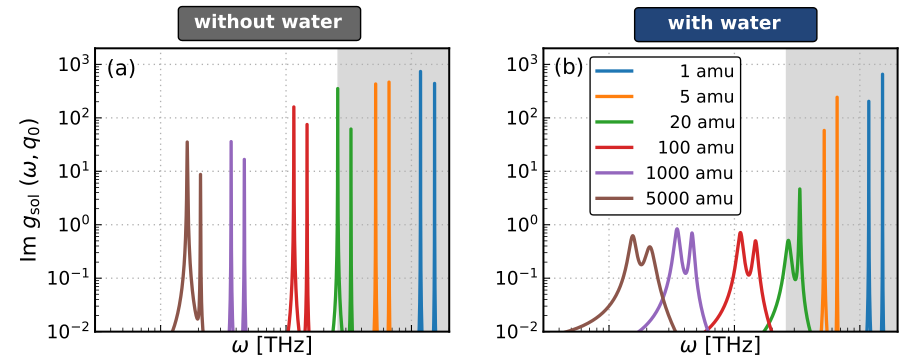}
    \caption{\textbf{Solid peaks broaden in the presence of water.}
    The solid surface response function fitted to a double-Lorentzian
    function for a range of different $m_{\mrm{D}}$ (as indicated in the legend)
    in the (a) absence and (b) presence of water.}
    \label{si:solid_width}
\end{figure}

In Figs.~\ref{si:solid_width}, we show the fitted $g_{\mrm{sol}}(q_0,\omega)$
profile for simulations of the solid with various values of $m_\mrm{D}$
in the absence and in the presence of water.
While the principal peaks remain at
$\omega_0\approx\omega_1\approx\omega_{\mrm{D}}$ in all cases, they
begin to broaden and decrease in height more significantly for cases
in the strong-coupling regime.

\subsection{Water surface response function}

\begin{figure}[H]
    \centering
    \includegraphics[width=\linewidth]{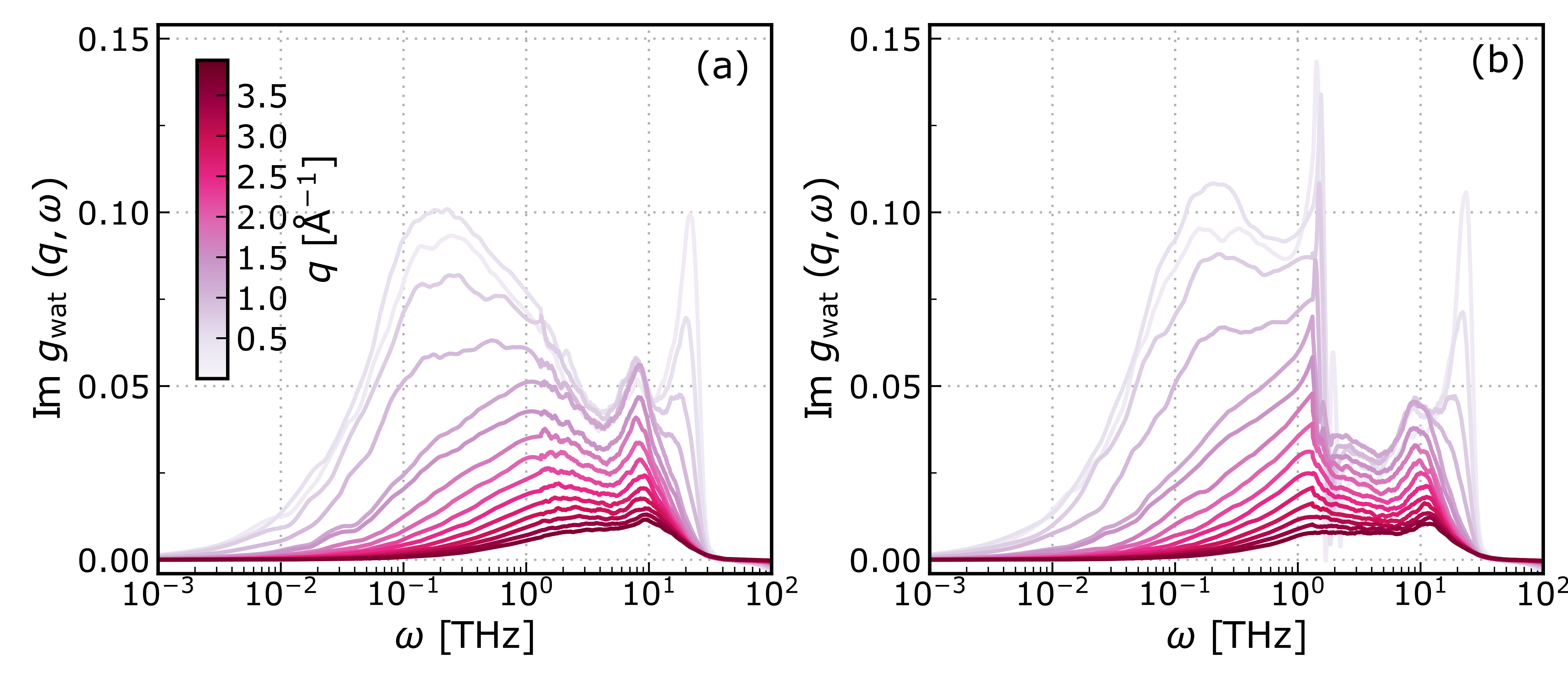}
    \caption{\textbf{Water surface response function.}
    $g_{\mrm{wat}}(q, \omega)$ obtained from simulations
    (a) in the weak-coupling regime appears unperturbed by the 
    presence of the Drude oscillators. (b) In the strong-coupling regime, $g_{\mrm{wat}}(q, \omega)$ is strongly
    perturbed by the solid up to $q\approx\SI{1.5}{\angstrom^{-1}}$.
    }
    \label{si:water_surface_response}
\end{figure}

For the water surface response in the weak-coupling regime, we focus on 
$g_{\mrm{wat}}(q, \omega)$ obtained from simulations with 
$m_{\mrm{D}}=\SI{1}{\amu}$ here as shown in Fig.~\ref{si:water_surface_response}(a).
However, the result pertains to all simulations 
with $m_{\mrm{D}}\lesssim\SI{20}{\amu}$.
At the long wavelength limit ($q \rightarrow 0$), we see that water shows
a sharp peak at $\omega_{\mrm{lib}}\approx\SI{20}{\THz}$ coming from the librational
modes and a broad feature spanning ~$10^{-2}-10^1\,\si{\THz}$.
The dielectric fluctuations due to intermolecular modes
of water in this regime agree well with previous simulations with
different interaction potentials \cite{Kavokine2022}.
As $q$ is increased, we see a decrease in $g_{\mrm{wat}}(q, \omega)$.
Most importantly, in the weak-coupling regime, the water
response function appears unperturbed by the presence of the Drude particles at all wavevectors.

In the strong-coupling regime, we focus on 
$g_{\mrm{wat}}(q, \omega)$ obtained from simulations with 
$m_{\mrm{D}}=\SI{5000}{\amu}$.
As shown in Fig.~\ref{si:water_surface_response}(b), at the long wavelength limit ($q \rightarrow 0$), 
we see that $g_{\mrm{wat}}(q, \omega)$ is strongly perturbed,
indicating that the water and the Drude particles' motions are strongly coupled.
As $q$ increases above $q\approx\SI{1.5}{\angstrom^{-1}}$, this coupling becomes less significant, which is consistent with $g_{\mrm{sol}}(q, \omega)$ decays for $q\gtrsim\SI{1.5}{\angstrom^{-1}}$.

\subsection{Force spectra}

To further illustrate the change from the weak-coupling 
to the strong-coupling regime, we also computed the spectrum of 
the lateral force defined as 
\begin{equation}
S_{\mrm{F}}(\omega)=\int^{+\infty}_{-\infty}\!
\mrm{d}t \,\langle\mcl{F}(0)\mcl{F}(t)\rangle \,
e^{i\omega t}.
\end{equation}

\begin{figure}[H]
    \centering
    \includegraphics[width=\linewidth]{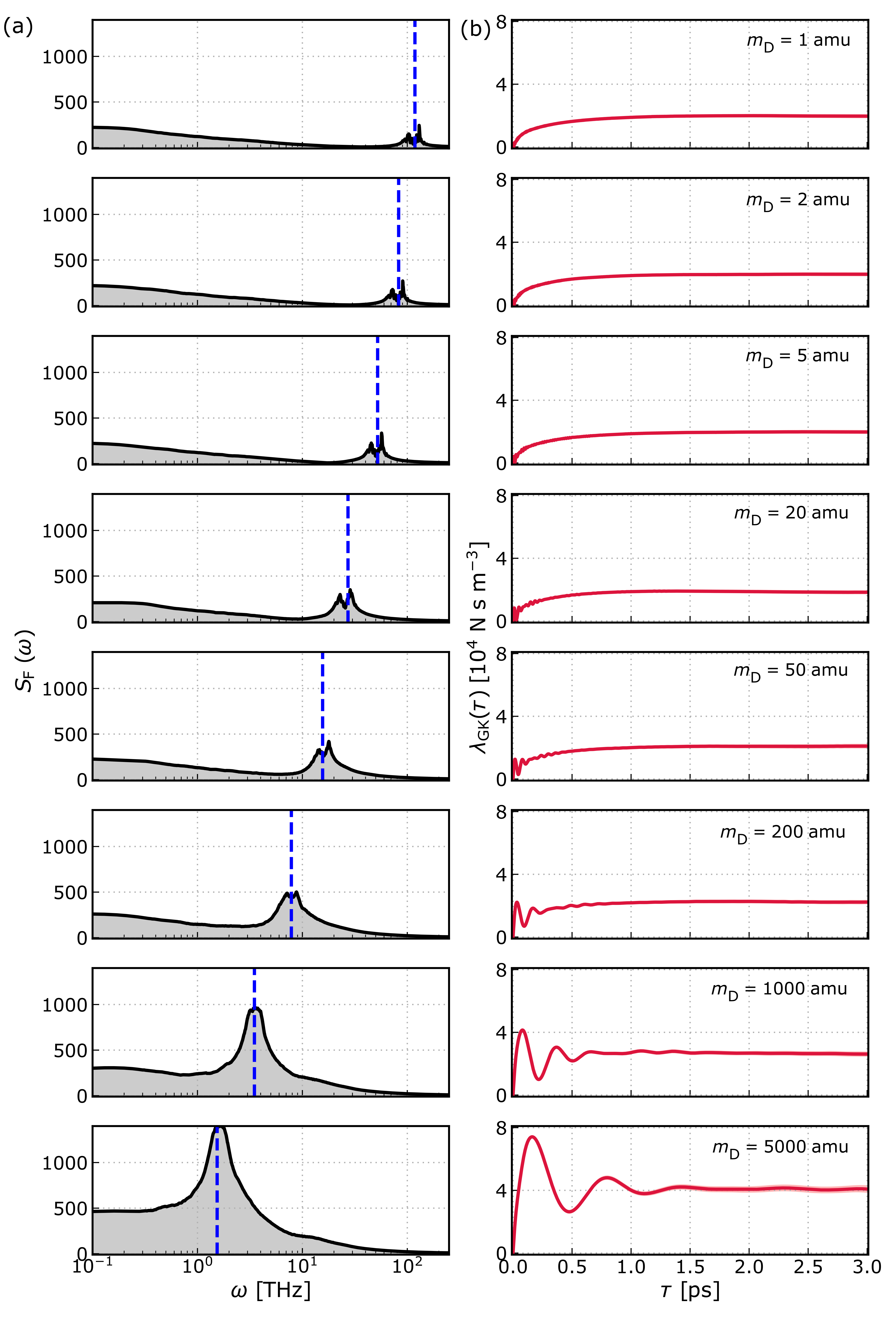}
    \caption{\textbf{Lateral force spectra.}
    (a) $S_{\mrm{F}}(\omega)$ for simulations with different $m_{\mrm{D}}$.
    The value of $\omega_{0}$ extracted from 
    $S_{\mrm{sol}}(q_0,\omega)$, marked as a dashed blue line,
    match with the frequency of the peak in $S_{\mrm{F}}(\omega)$ 
    due to solid charge density contribution. This peak 
    shows higher intensity as we move from the weak-coupling regime
    ($m_{\mrm{D}}<20\,\mrm{amu}$) to the strong-coupling regime
    ($m_{\mrm{D}}\geq20\,\mrm{amu}$). (b) This change manifests
    in $\lambda_{\mrm{GK}}(\tau)$ as the appearance of small oscillations
    at short $\tau$ and the increase in the plateau value at longer $\tau$.}
    \label{si:force_spectra}
\end{figure}

We show $S_{\mrm{F}}(\omega)$ for simulations with Drude 
masses used the previous subsection 
in conjunction with the Green--Kubo friction 
$\lambda_{\mrm{GK}}(\tau)$ in Fig.~\ref{si:force_spectra}.
For cases belonging to the weak-coupling regime 
($m_{\mrm{D}}/\mrm{amu}=1, 2,\,\mrm{and}\,5$),
$S_{\mrm{F}}(\omega)$ shows a peak due to the
solid charge density at $\omega\approx\omega_{0}$
determined from $S_{\mrm{sol}}(q_0,\omega)$,
and a broad feature at low $\omega$ from the water charge density.
As $m_{\mrm{D}}$ is increased, the peak due to the solid
starts to merge with the water broad feature and increase
in intensity.
This can be linked to the behavior of $\lambda_{\mrm{GK}}(\tau)$
in the strong-coupling regime ($m_{\mrm{D}}/\mrm{amu}=20, 50, 200, 1000\,\mrm{and}\,5000$):
the increase in $S_{\mrm{F}}(\omega=\omega_{0})$ is responsible for
stronger oscillations in $\lambda_{\mrm{GK}}(\tau)$ at shorter times ($\tau<\tau_\mrm{F}$)
while the increase in $S_{\mrm{F}}(\omega\lesssim\omega_{0})$
is responsible for a higher plateau value at longer times ($\tau \geq\tau_\mrm{F}$).

\newpage

\section{Comparison to quantum friction theory}

\subsection{Quantum friction formula}

In quantum friction theory\cite{Kavokine2022}, Kavokine \etal{} derived an expression for 
the QF coefficient given as
\begin{equation}
  \lambda_{\rm Q} = \frac{\hbar^{2}}{8\pi^{2}k_{\rm B}T}
  \int^{\infty}_{0}\!\mrm{d}q\, q^3
  \int^{\infty}_{0}\! \mrm{d}\omega\, \frac{1}{\sinh^2 (\hbar\omega/2k_{\rm B}T)}
  \frac{\mrm{Im}\,g_{\rm sol}(q,\omega)\,\mrm{Im}\,g_{\rm wat}(q,\omega)}{|1- g_{\rm sol}(q,\omega)\,g_{\rm wat}(q,\omega)|^2},
  \label{eqn:QF}
\end{equation}
where $g_{\rm sol}(q,\omega)$ and $g_{\rm wat}(q,\omega)$ are the surface response
functions for the solid and water in the absence of any coupling respectively.
%

\subsection{Water surface response}

To evaluate $\lambda_{\rm Q}$ quantitatively, one needs to obtain $g_{\rm sol}(q,\omega)$ 
and $g_{\rm wat}(q,\omega)$. 
As in Ref.~\onlinecite{Kavokine2022}, the water dielectric function is represented by
a sum of two Debye peaks at $\omega_{\mrm{Db},1}$ and $\omega_{\mrm{Db},2}$,
each with an exponentially decaying $q$ dependence
\begin{equation}
  g_{\rm wat}(q,\omega) = \frac{g_{\rm wat}(q,0)}{2}
  \left( \frac{\mrm{e}^{-q/q_0}}{1-i\omega/\omega_{\mrm{Db},1}} 
  + \frac{2-\mrm{e}^{-q/q_0}}{1-i\omega/\omega_{\mrm{Db},2}} \right),
\end{equation}
where $g_{\rm wat}(q,\omega)$ is given as
\begin{equation}
g_{\rm wat}(q,0) = \mrm{e}^{a + b [1+ (q/c)^{d}]^{1/d} }.
\end{equation}
We have used the same parameters as Ref.~\onlinecite{Kavokine2022}, which for completeness,
are reproduced in Table~\ref{QFwaterparameters}.

\begin{table}[H]
    \centering
    \begin{tabular}{>{\color{black}}l >{\color{black}}l >{\color{black}}l}
    \hline
    \hline
     $g_{\rm wat}(q,\omega)$  &  Werder $g_{\rm wat}(q,0)$  & Aluru $g_{\rm wat}(q,0)$\\
    \hline
     $q_0 = 3.12\,\angstrom^{-1}$    & $a=5.16$  & $a=3.38$\\
     $\omega_{\mrm{Db},1}=\SI{0.36}{\THz}$    & $b=-5.19$  & $b=-3.41$\\
     $\omega_{\mrm{Db},2}=\SI{4.84}{\THz}$   & $c=1.95\,\angstrom^{-1}$ & $c=1.79\,\angstrom^{-1}$\\
                                                & $d = 2$ & $d = 2.4$\\
    \hline
    \hline
\end{tabular}
\caption
{\textbf{Parameters for the analytical expression for the water surface response function
in the absence of charge density coupling from the solid.}}
\label{QFwaterparameters}
\end{table}

For the solid dielectric fluctuations, two different models were used to represent
the graphite's dispersionless surface plasmon.
Numerical evaluation in Ref.~\onlinecite{Kavokine2022} of the QF
coefficient using the Aluru $g_{\rm wat}(q,\omega)$ gave a
contribution of $\lambda_{\mrm{Q}}\approx 0.5\times 10^{4}\,\si{\Nsm}$
with a ``Drude'' model for the surface plasmon, which is comparable to
$\lambda_{\rm THz}$ obtained from our simulations.

%

\subsection{Solid surface response with reparameterized Drude model}

Encouraged by the agreement between $\lambda_{\rm THz}$ obtained from
simulations and $\lambda_{\rm Q}$ predicted from QF theory, we can
further assess how well our simulations are capturing QF by
reparameterizing the Drude model to approximately represent the
surface response function of our simple model.
In the Drude model for a surface plasmon, which is based on the
semi-classical treatment of free electron dynamics \cite{Pitarke2006},
the solid surface response function is of the form
\begin{equation}
g_{\rm sol}(q,\omega) = \frac{\omega_{\rm p}^2}{\omega_{\rm p}^2 - \omega^2 - 2i\eta\,\omega}\Theta(q_{\rm max}-q),
\end{equation}
where $\omega_{\rm p}$ is the principal peak due to the plasmon, $\eta$ is the
surface plasmon width, $\Theta$ is the Heaviside step function and $q_{\rm max}$
is the cut-off wavevector.

Drawing a parallel mapping to the surface response function obtained
in our simulations, we can parameterize the Drude model to represent a
surface plasmon with frequency $\omega_{\rm p}=\omega_{0}$
corresponding to the principal tangential mode in simulations. We
choose $\eta=\omega_{0}/100$ and $q_{\rm
max}=\SI{1.5}{\angstrom^{-1}}$, giving $\mrm{Im}\,g_{\rm
sol}(q<\SI{1.5}{\angstrom^{-1}}, \omega=\omega_{\mrm{p}})=50$, such
that the plasmon has a small width and long flat dispersion, and its
surface response function matches reasonably well with that obtained
in simulations (Fig.~\ref{si:width_dispersion}).

\subsection{Comparison of the quantum friction coefficient} 

\begin{figure}[H]
    \centering \includegraphics[width=\linewidth]{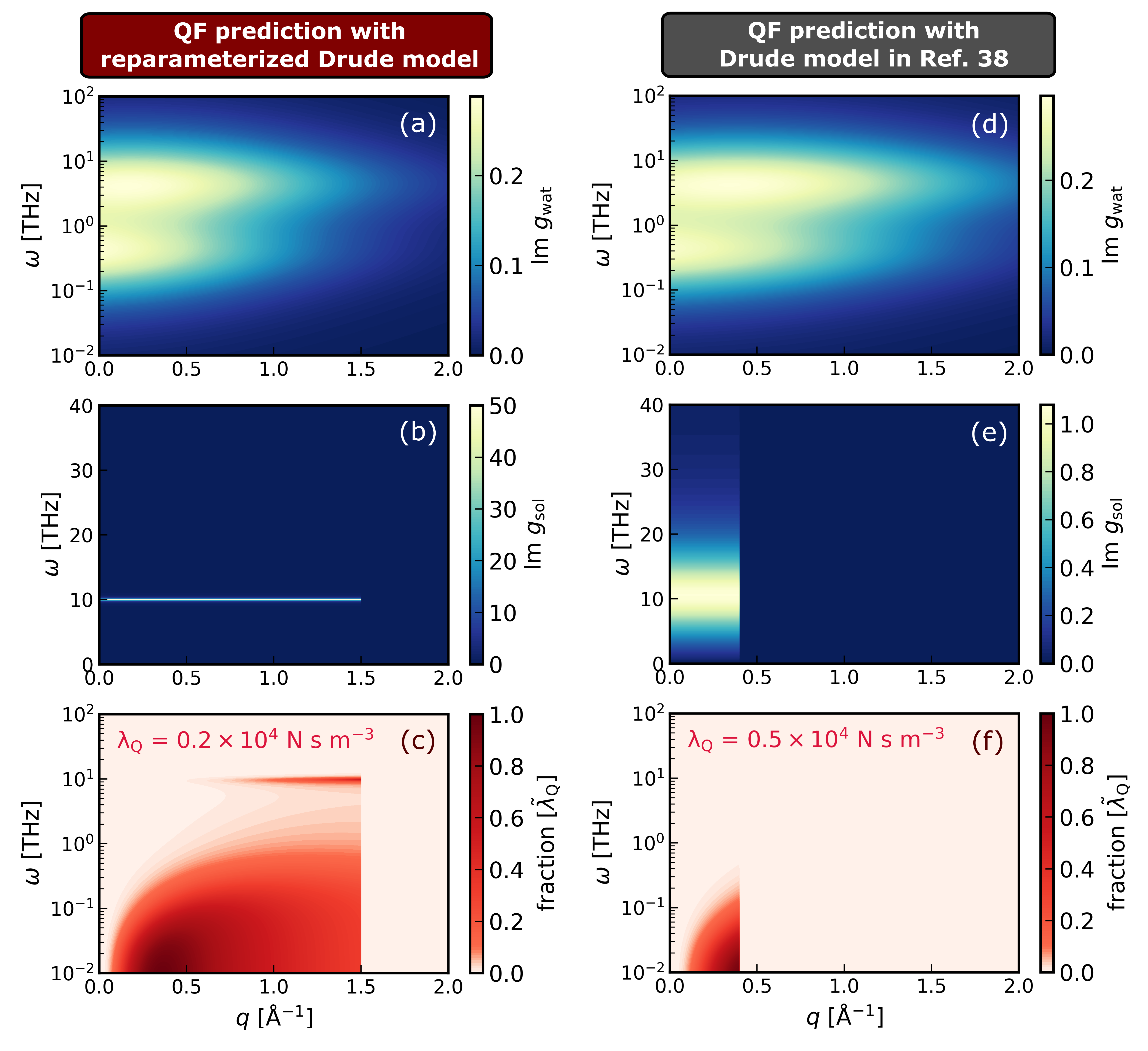} \caption{\textbf{Comparison
    of the quantum friction coefficient with parameters from
    simulations (a-c) and from Ref.~\onlinecite{Kavokine2022} (d-f).}
    (a) The water Werder surface response function.  (b) The solid
    surface response function from the reparameterized Drude model.
    (c) The QF integrand contributes mainly at
    ($\omega\lesssim\SI{1}{\THz}$, $q<\SI{1.5}{\angstrom^{-1}}$) and
    at ($\omega\approx\omega_{\mrm{p}}=\SI{10}{\THz}$,
    $1<q/\angstrom^{-1}<1.5$).  (d) The water Aluru surface response
    function.  (e) The solid surface response function from the Drude
    model with parameters from Ref.~\onlinecite{Kavokine2022}.
    (f) The QF integrand contributes mainly at
    ($\omega\lesssim\SI{0.1}{\THz}$, $q<\SI{0.5}{\angstrom^{-1}}$).
    } \label{si:compare_QF_decompose}
\end{figure}

Using the reparameterized Drude model for the solid surface response 
function and the Werder water surface response function, we evaluated 
the QF coefficient $\lambda_{\mrm{Q}}$ from Eq.~\ref{eqn:QF}. 
For a plasmon mode with $\omega_{\mrm{p}}=\SI{10}{\THz}$, 
the reparameterized Drude model gives $\lambda_{\mrm{Q}}\approx 0.2\times 10^{4}\,\si{\Nsm}$.
This is close to the value obtained with the Drude model in
Ref.~\onlinecite{Kavokine2022} and the quantitative difference can be
explained by comparing the contribution of the integrand of
Eq.~\ref{eqn:QF}, $\tilde{\lambda}_{\mrm{Q}}$, in $(q,\omega)$ space,
as shown in Fig.~\ref{si:compare_QF_decompose}.

Since there is little difference between the water surface response
functions $g_{\mrm{wat}}(q,\omega)$ in the two cases, the difference
comes down to the solid surface response $g_{\mrm{sol}}(q,\omega)$ 
or ultimately the plasmon dispersion. 
Since the plasmon in the reparameterized Drude model spans to a higher
wavevector compared to the Drude model in
Ref.~\onlinecite{Kavokine2022}, $\tilde{\lambda}_{\mrm{Q}}$
contributes significantly to the final integral not only at
($\omega\lesssim\SI{1}{\THz}$, $q<\SI{1.5}{\angstrom^{-1}}$) but also
at ($\omega\approx\omega_{\mrm{p}}=\SI{10}{\THz}$,
$1<q/\angstrom^{-1}<1.5$).
However, since the width of the plasmon in the reparameterized Drude model
is much smaller, the final integral value for $\lambda_{\mrm{Q}}$ is still smaller.

\subsection{Sensitivity of the dependence of quantum friction
on the solid frequency}

From QF theory, we can also map out the dependence of the QF friction
coefficient as a function of frequency of the solid mode.
Using the reparameterized Drude model, as presented in the main
article, we see a very good agreement between $\lambda_{\mrm{Q}}$
calculated from theory and  $\lambda_{\mrm{THz}}$ obtained from simulations
in the frequency range where graphite surface plasmons are 
experimentally observed.

\begin{figure}[H]
    \centering
    \includegraphics[width=\linewidth]{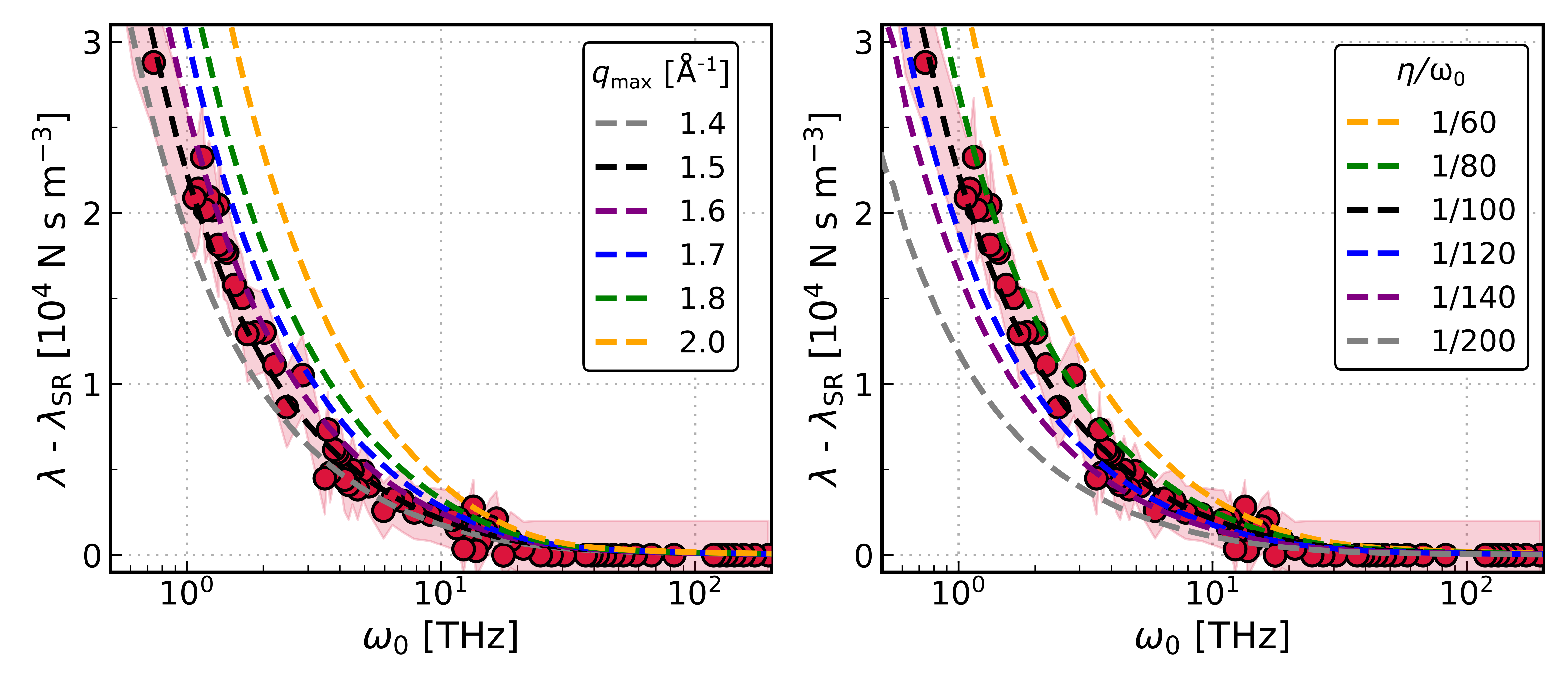}
    \caption{
    \textbf{Dependence of the quantum friction coefficient
    on the solid charge density frequency from the Drude model} (a) for 
    different values of $q_{\mrm{max}}$ (indicated in the legend) while 
    keeping $\eta=\omega_0/100$ and (b) for different values of 
    $\eta/\omega$ keeping $q_{\mrm{max}}=\SI{1.5}{\angstrom^{-1}}$. The
    calculated $\lambda_{\mrm{Q}}$ from theory is shown with
    dashed lines while $\lambda_{\mrm{THz}}$ extracted from simulations
    are shown as red points with the shaded red region as the error
    from block-averaging.}
    \label{si:QFsensitivity}
\end{figure}

In the Drude model, both the width $\eta$ and the wavevector
cut-off $q_{\mathrm{max}}$ control the plasmon dispersion and changing 
their values will affect the solid surface response function
$g_{\mrm{sol}}(q,\omega)$ and therefore the friction. 
We can also check the sensitivity of the dependence of the
QF coefficient on the plasmon frequency upon changing to 
different values for each of these parameters.
In simulations with the solid modes' frequencies in the range
$2-20\,\si{\THz}$ where graphite's surface plasmons are experimentally
observed, the solid modes' wavevector cut-offs can sensibly range
between $q_{\mrm{max}}\approx1.4-2.0\,\si{\angstrom^{-1}}$ and their
amplitudes can sensibly range between
$\mrm{Im}\,g_{\mrm{sol}}(q<q_{\mrm{max}}, \omega=\omega_{\mrm{p}})\approx
30-100$. This corresponds to $\eta/\omega_{\mrm{0}}\approx1/60-1/200$
in the Drude model.
In Fig.~\ref{si:QFsensitivity}(a), we show the dependence
of QF on $\omega_0$ using the Drude model for different values
of $q_{\mrm{max}}$ while keeping the width at $\eta=\omega_0/100$.
As $q_{\mrm{max}}$ increases, there is a higher contribution to the
QF integral at higher $q$, leading to a higher $\lambda_{\mrm{Q}}$.
In Fig.~\ref{si:QFsensitivity}(b), we show again this dependence
for different values of the width $\eta/\omega_0$ while 
while keeping the wavevector cut-off at $q_{\mrm{max}}=\SI{1.5}{\angstrom^{-1}}$.
As $\eta/\omega_0$ increases, the width of the plasmon decreases,
leading to lower values of $\lambda_{\mrm{Q}}$.
In all cases, however, there is only a significant contribution
to QF when $\omega_{0}\lesssim\SI{20}{\THz}$, which agrees with 
the results in the strong-coupling regime from simulations.

\newpage
\section{Additional properties of the interface}
\subsection{Static properties}

\begin{figure}[H]
    \centering
    \includegraphics[width=\linewidth]{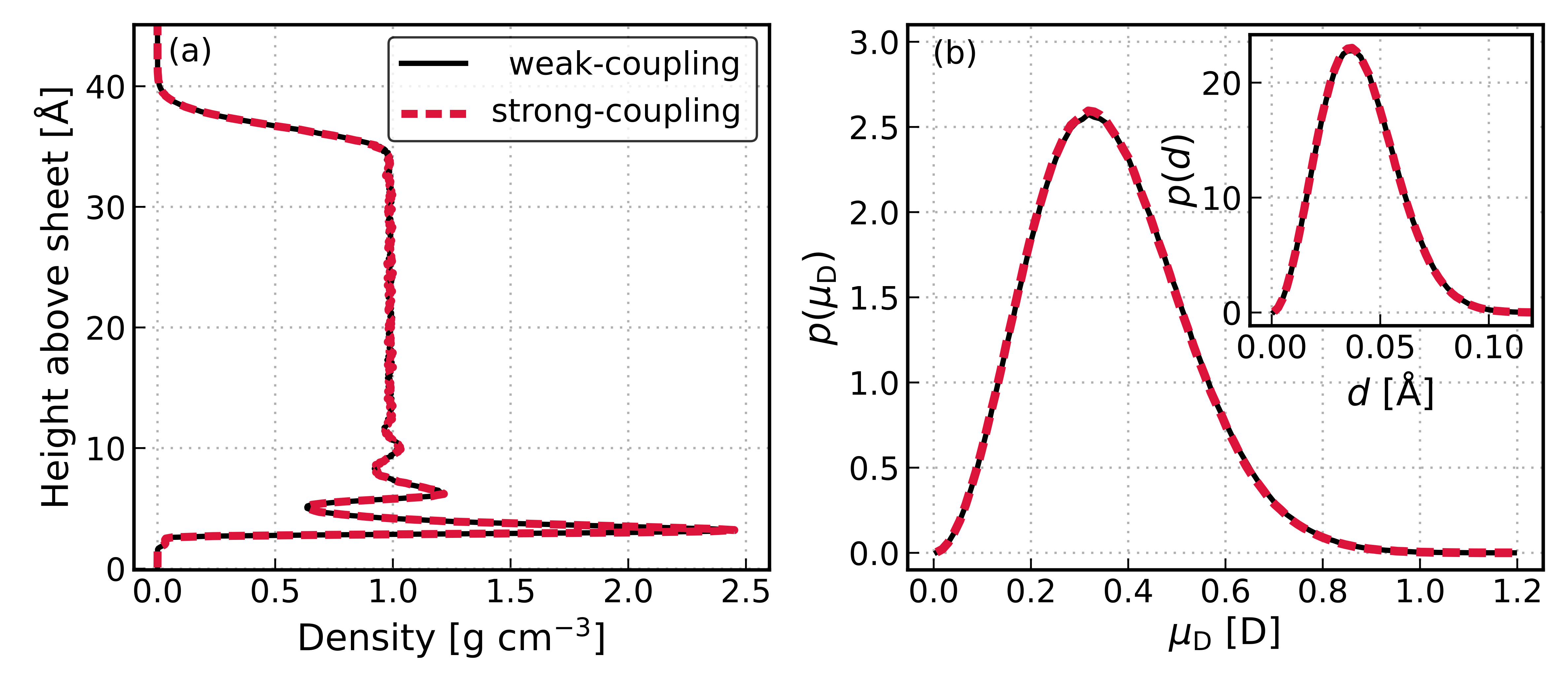}
    \caption{\textbf{Static equilibrium properties.}
    (a) The planar mass density profiles of the water are
    identical for the weak-coupling and the strong-coupling
    cases, showing a maximum density of 
    $\approx \SI{2.5}{\gram\per\centi\meter\cubed}$
    and is at a height $\approx 3.2\,\angstrom$ for the
    contact layer of water. (b) The probability distributions 
    of the dipole moment magnitude of a Drude oscillator
    $p(\mu_{\mrm{D}})$ in the solid are also identical for both cases,
    with the average of the distribution at 
    $\langle\mu_{\mrm{D}}\rangle\approx 0.36\,\mrm{D}$.
    The inset shows the distribution of the distance between the Drude
    particles and their cores $p(d)$.
    }
    \label{si:staticproperties}
\end{figure}

For static properties in the liquid, we analyse the density 
profiles of the water along the surface normal.
These are identical for the weak-coupling and the strong-coupling 
cases, as shown in Fig.~\ref{si:staticproperties}(a).
In the solid, the magnitude of the dipole moment of
a Drude oscillator can be obtained from 
$\mu_{\mrm{D}}=Q_{\mrm{D}}d$ where $d$ is the distance
of the Drude particle from the core atom.
We look at its probability distribution $p(\mu_{D})$
for both cases, as shown in Fig.~\ref{si:staticproperties}(b).
In both cases, $p(\mu_{D})$ is identical, reinforcing that 
the interatomic potential, and therefore 
all static equilibrium properties of the 
interface are not affected when $m_{\mrm{D}}$ is varied.

\begin{figure}[H]
    \centering
    \includegraphics[width=0.95\linewidth]{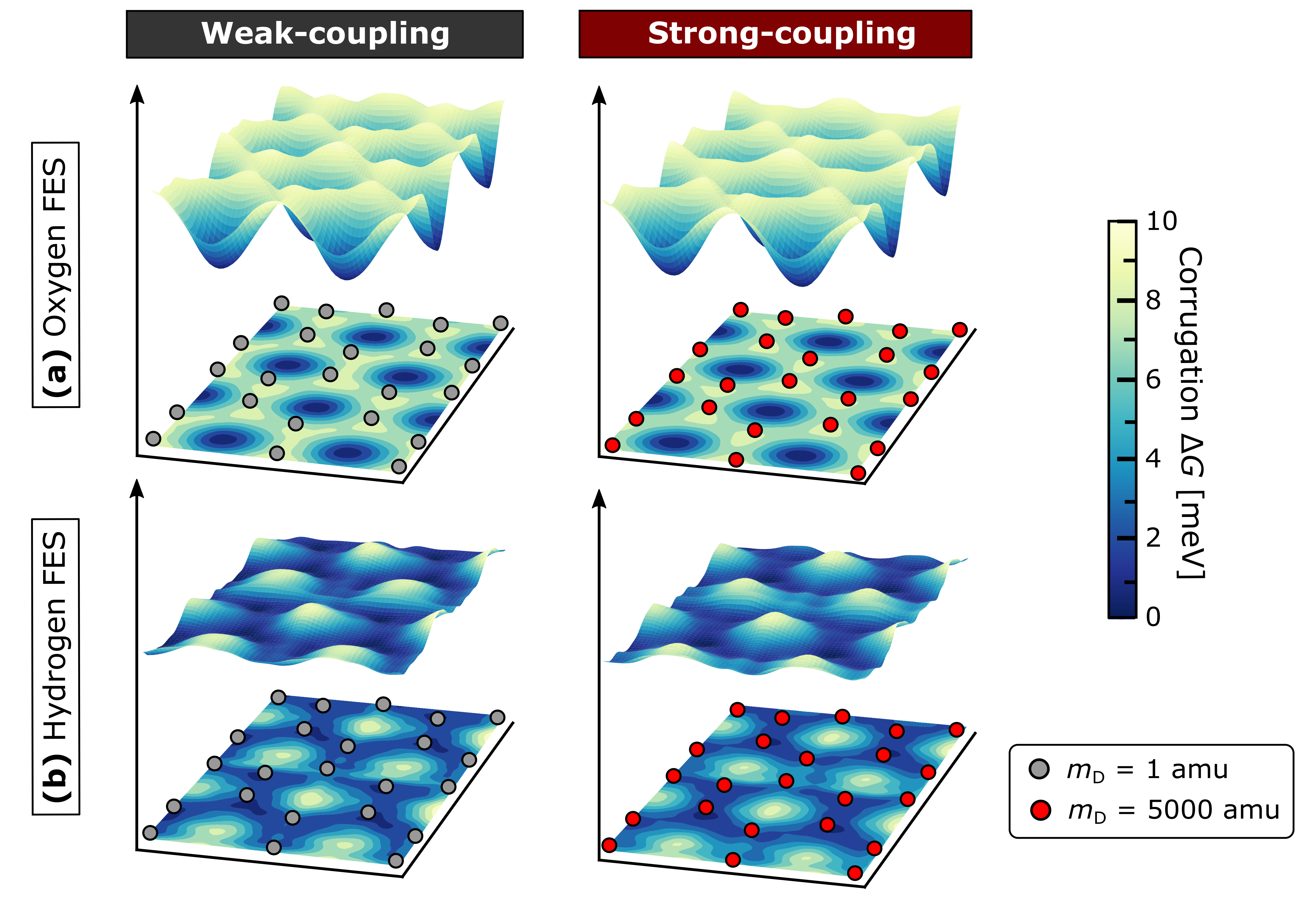}
    \caption{\textbf{Free energy surface corrugation}
    (a) The oxygen-based FES shows that oxygen atoms preferentially
    sit on hollow sites in the middle of the hexagon rings in 
    graphene. (b) The hydrogen-based FES, which is less corrugated than
    the oxygen-based FES, shows that hydrogen atoms preferentially sit
    on carbon sites.
    The solid atoms are represented by the markers in the projection
    where the core atom in the Drude oscillator with
    $m_{\mrm{D}}=1\,\mrm{amu}$ is in grey and $m_{\mrm{D}}=5000\,\mrm{amu}$ 
    in red.
    Both FESs show identical level of corrugation 
    in the weak-coupling and strong-coupling cases.
    }
    \label{si:fes}
\end{figure}

To link these observations to the unchanged static component 
of the friction coefficient described in the main article, 
we also look at the corrugation of the free energy surface (FES)
experienced by the water molecules.
Following previous work \cite{Tocci2014,Tocci2020,Thiemann2022},
the two-dimensional FES of species $i$ is given by
\begin{equation}
  \Delta G_{i}(x,y) = -k_{\mrm{B}}T\,\mrm{ln}[p_{i}(x,y)], 
\end{equation}
where $p_{i}(x,y)$ is the normalized two-dimensional probability
of finding species $i$ in the contact layer at point $(x,y)$.
For every saved configuration, we define the contact layer
as consist of water molecules with height above the sheet 
$\leq 5\,\angstrom$, where the first minimum in the density 
profile is.
After computing both $p_{\mrm{O}}(x,y)$ and $p_{\mrm{H}}(x,y)$ and
averaging onto a unit cell in the solid, we obtained 
the oxygen-based and hydrogen-based FESs.
To ensure each surface is independent of noise, a
Savitzky--Golay filter \cite{Savitzky1964}
was applied. 
The corrugation of each FES, $\Delta G_{\mrm{O}}$ and
$\Delta G_{\mrm{H}}$,
can be quantified by taking the highest free energy
present in the FES.
As shown in Fig.~\ref{si:fes}, the water molecules
experience the same oxygen-based and hydrogen-based
FES at the liquid--solid interface in the
weak-coupling and the strong-coupling regimes.
Since corrugation is much more pronounced in
the oxygen-based FES, we show $\Delta G_{\mrm{O}}$
as an approximation of the total corrugation
in the main article.

\subsection{Charge density relaxation in the water film}

\begin{figure}[H]
    \centering
    \includegraphics[width=\linewidth]{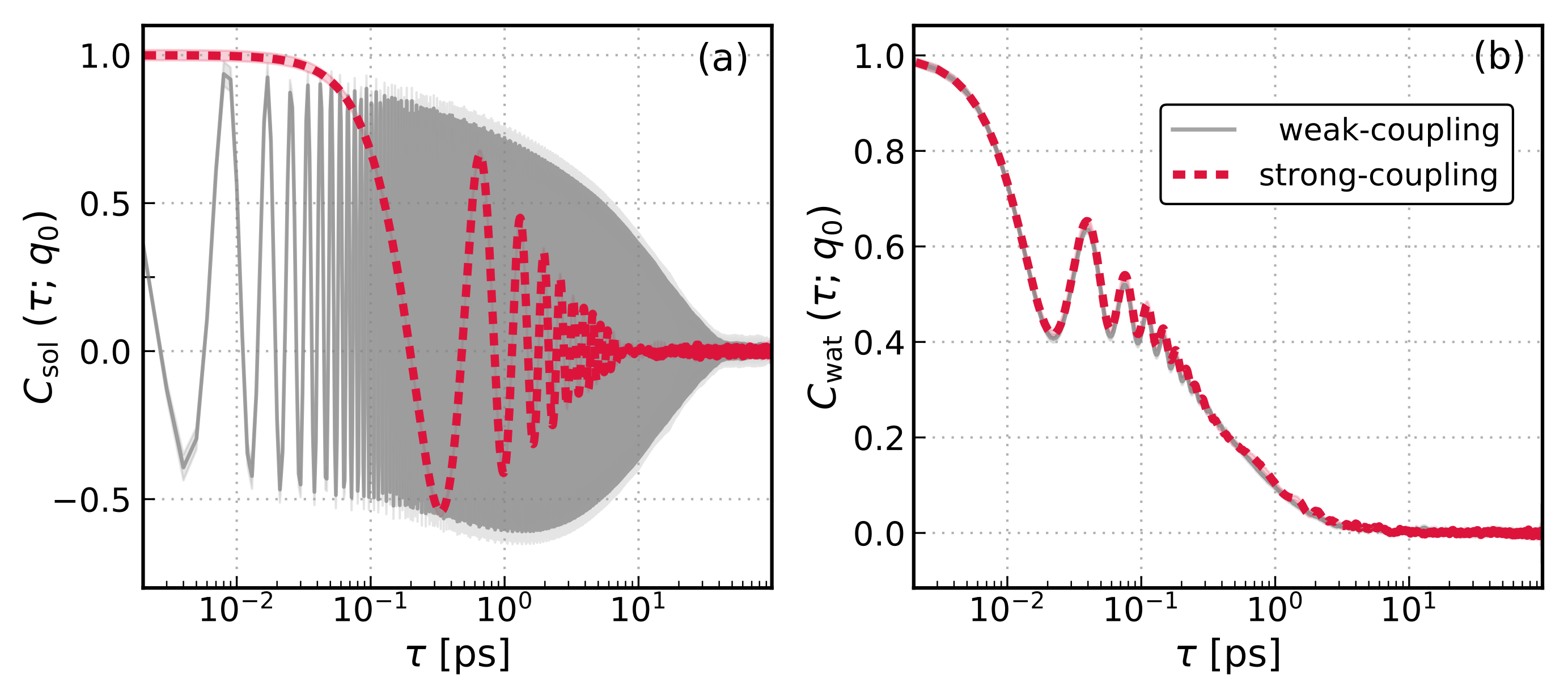}
    \caption{\textbf{Charge density correlation: asymmetry in response between 
    the liquid and the solid.} (a) The relaxation of the solid charge density
    is much faster in the strong-coupling than the weak-coupling regime while
    (b) relaxation of the charge density of the whole water film barely
    differs between the two regimes.
    }
    \label{si:wholeslab}
\end{figure}

Instead of looking at the charge density relaxation at just the
surface, we can also characterize the relaxation of the whole water
film.
We can define the Fourier components of the charge densities 
for the solid and the liquid as
\begin{equation}
\tilde{n}_{\mrm{sol}}(q,t) = \sum_{\alpha\,\in\,\mrm{sol}}
Q_{\alpha}e^{i\mbf{q}\cdot\mbf{x}_{\alpha}(t)},
\end{equation}
\begin{equation}
\tilde{n}_{\mrm{wat}}(q,t) = \sum_{\alpha\,\in\,\mrm{wat}}
Q_{\alpha}e^{i\mbf{q}\cdot\mbf{x}_{\alpha}(t)},
\end{equation}
where we have implicitly only considered the zero wavevector 
in the direction normal to the graphene sheet.
We can again characterize the relaxation of these charge densities
with the following autocorrelation functions
\begin{equation}
  C_{\mrm{sol}}(\tau; q) = \frac{\langle\tilde{n}_{\mrm{sol}}(q,0)
  \,\tilde{n}_{\mrm{sol}}(-q,\tau)\rangle}
  {\langle|\tilde{n}_{\mrm{sol}}(q)|^2\rangle},
\end{equation}
\begin{equation}
  C_{\mrm{wat}}(\tau; q) = \frac{\langle\tilde{n}_{\mrm{wat}}(q,0)
  \,\tilde{n}_{\mrm{wat}}(-q,\tau)\rangle}
  {\langle|\tilde{n}_{\mrm{wat}}(q)|^2\rangle}.
\end{equation}
Focusing on the long-wavelength limit, we
show the results for $C_{\mrm{sol}}(\tau; q_0)$
and $C_{\mrm{wat}}(\tau; q_0)$ in Fig.~\ref{si:wholeslab}.
Again, the solid modes relax on a much faster timescale in the strong-
than in the weak-coupling regime.
The water relaxation, however, barely differs between the two regimes,
meaning any response in the liquid due to coupling with the Drude
motions is localized the surface and does not affect the response of
the whole film of water significantly as a whole.

\subsection{Other dynamical properties}

In addition to the charge density relaxation, we also
explored other dynamical properties of water 
including its orientational relaxation and hydrogen-bonding
relaxation.
The orientational dynamics of water molecules in the liquid
is examined, following Yeh and Mou \cite{Yeh1999}, via the
second-order rotational autocorrelation function, defined as 
\begin{equation}
  C_{\mrm{rot}}(\tau) = \langle P_2 [\mbf{u}(0)
  \cdot\mbf{u}(\tau)] \rangle ,
\end{equation}
where $\mbf{u}(\tau)$ is the unit vector along the water
molecular dipole at time $\tau$ and $P_{2}(x)$ denotes the
second Legendre Polynomial.
The hydrogen-bond relaxation is examined via the autocorrelation 
function of the presence of a hydrogen bond, defined as
\begin{equation}
  C_{\mrm{hb}}(\tau) = \frac{\langle h(0)
  \cdot h(\tau)\rangle}
  {\langle h^2\rangle},
\end{equation}
where $h(\tau)=1$ if there is a hydrogen bond between 
a pair of water molecules at time $\tau$ and $h(\tau)=0$ otherwise.
Two water molecules are considered to be hydrogen-bonded 
according to geometric criteria from Luzar and Chandler \cite{Luzar1996}.

\begin{figure}[H]
    \centering
    \includegraphics[width=\linewidth]{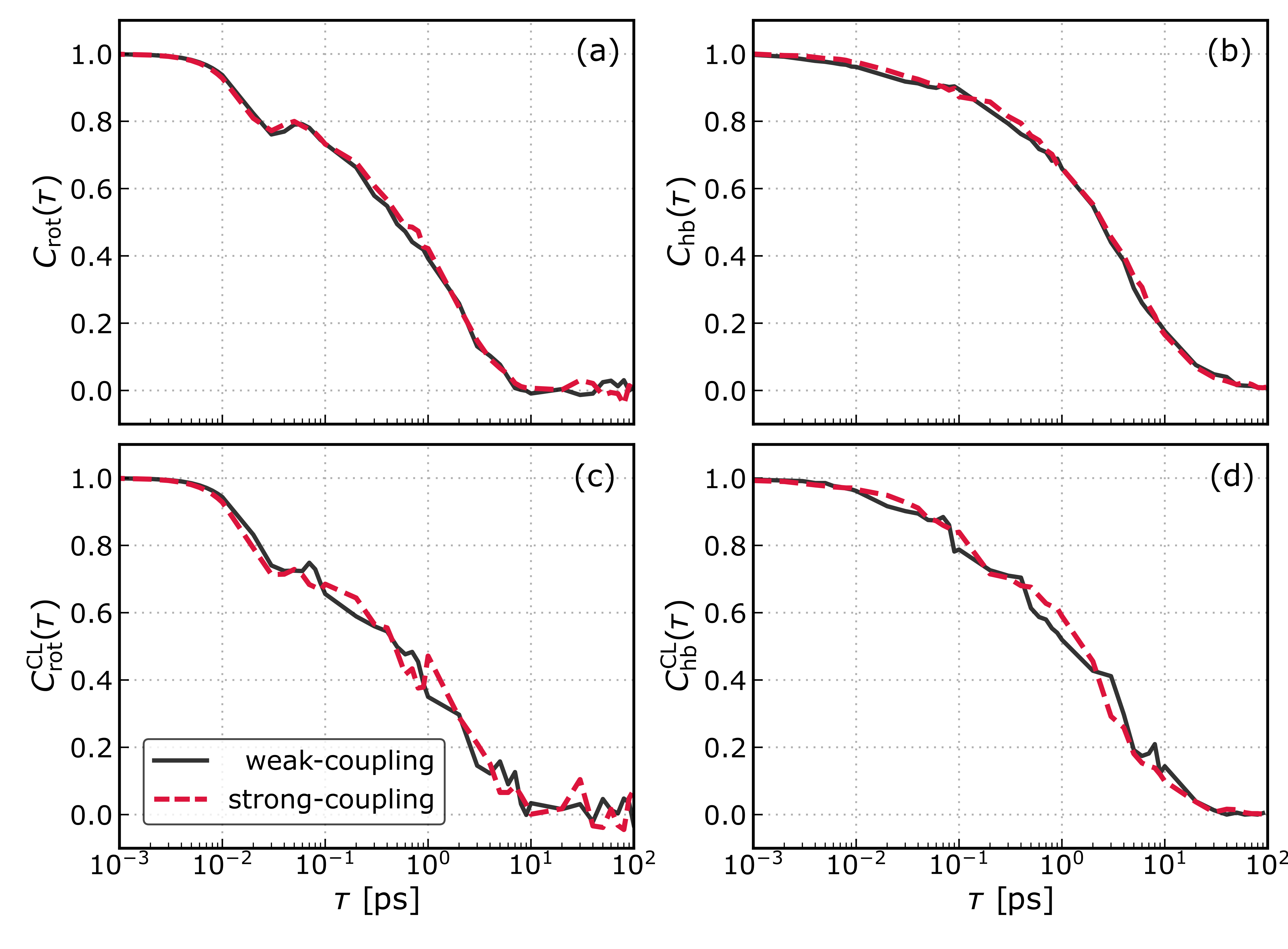}
    \caption{\textbf{Dynamical properties of water.} 
    The rotational autocorrelation (a and c) and the hydrogen-bonding 
    autocorrelation (b and d) are both barely affected between the weak-coupling
    and the strong-coupling regimes. The functions are computed for
    the whole water film (a and b) and just the contact layer (c and d).
    }
    \label{si:water_dynamics}
\end{figure}

From Figs.~\ref{si:water_dynamics}(a) and (b), we see little differences between the
weak-coupling and strong-coupling cases for both $C_{\mrm{rot}}(\tau)$
and $C_{\mrm{hb}}(\tau)$ computed for the whole water film.
These observations still hold when we look at $C_{\mrm{rot}}^{\mrm{CL}}(\tau)$
and $C_{\mrm{hb}}^{\mrm{CL}}(\tau)$, in Figs.~\ref{si:water_dynamics}(c) and (d),
where superscript $\mrm{CL}$ denotes that the autocorrelation 
functions are defined for just the contact layer (defined as the layer
from the carbon sheet up to the first minimum of the water density profile).
This supports our conclusion that the increase in friction due to
charge density coupling has little impact on local dynamical
properties of the liquid.

\newpage
\bibliography{si_bibliography.bib}